\documentclass[graphicx]{ptephy_v1}

\usepackage{graphicx}
\usepackage{dcolumn}
\usepackage{bm}
\usepackage{comment}
\usepackage{setspace}
\usepackage{subfigure}
\usepackage{here}
\usepackage{lscape}
\usepackage{tabularx}
\usepackage{array}
\usepackage{threeparttable}

\usepackage{version}	
\usepackage{amsmath}	

\newcolumntype{C}{>{\centering\arraybackslash}X}
\newcolumntype{R}{>{\raggedright\arraybackslash}X}
\newcolumntype{L}{>{\raggedleft\arraybackslash}X}
\makeatletter
\newcommand{\figcaption}[1]{\def\@captype{figure}\caption{#1}}
\newcommand{\tblcaption}[1]{\def\@captype{table}\caption{#1}}
\makeatother

\begin{document}

\title{Search for $\alpha$ condensed states in $^{13}$C using $\alpha$ inelastic scattering}


\author{K.~Inaba}
\affil{Department of Physics, Kyoto University, Sakyo, Kyoto 606-8502, Japan \email{kento@nh.scphys.kyoto-u.ac.jp}}

\author[2,15]{Y.~Sasamoto}
\affil{Center for Nuclear Study, The University of Tokyo, Bunkyo, Tokyo 113-0033, Japan}

\author{T.~Kawabata}
\affil{Department of Physics, Osaka University, Toyonaka, Osaka 560-0043, Japan}

\author{M.~Fujiwara}
\affil{Research Center for Nuclear Physics, Ibaraki, Osaka University, Osaka 567-0047, Japan}
\author{Y.~Funaki}
\affil{College of Science and Engineering, Kanto Gakuin University, Yokohama, Kanagawa 236-8501, Japan}
\author[4]{K.~Hatanaka} 
\author{K.~Itoh}%
\affil{Department of Physics, Saitama University, Sakura, Saitama 338-8570, Japan}
\author{M.~Itoh}%
\affil{Cyclotron and Radioisotope Center, Tohoku University, Sendai, Miyagi 980-8578, Japan}
\author{K.~Kawase}%
\affil{National Institutes for Quantum and Radiological Science and Technology, Tokai, Ibaragi 319-1106, Japan}
\author[4,16]{H.~Matsubara}%
\author{Y.~Maeda}%
\affil{Faculty of Engineering, University of Miyazaki, Miyazaki 889-2192, Japan}
\author{K.~Suda}%
\affil{RIKEN Nishina Center for Accelerator-Based Science, Wako, Saitama 351-0198, Japan}
\author{S.~Sakaguchi}%
\affil{Department of Physics, Kyushu University, Fukuoka 819-0395, Japan} 
\author[10]{Y.~Shimizu}%
\author[3,4,12]{A.~Tamii}%
\affil{Institute for Radiation Science, Osaka University, Ibaraki, Osaka 560-0043, Japan}
\author{Y.~Tameshige}%
\affil{Fukui Prefectural Hospital, Fukui 910-8526, Japan}
\author{M.~Uchida}%
\affil{Department of Physics, Tokyo Institute of Technology, Ota, Tokyo 152-8551, Japan}
\author[10]{T.~Uesaka}%
\author[5]{T. Yamada}
\affil{Present address: Research Promotion Office, Chuo University, Bunkyo, Tokyo 112-8551, Japan}
\author[4]{H.~P.~Yoshida}%
\affil{Present address: Department of Radiography, Kyoto Prefectural University of Medicine, Kamigyo, Kyoto 602-8566, Japan}



\begin{abstract}%
We searched for the $\alpha$ condensed state in $^{13}$C by measuring the $\alpha$ inelastic scattering at $E_\alpha = 388$ MeV at forward angles including 0 degrees.
We performed the distorted-wave Born-approximation calculation with the single-folding potential and the multipole decomposition analysis to determine the isoscalar transition strengths in $^{13}$C.
We found a bump structure around $E_x = 12.5$ MeV due to the isoscalar monopole ($IS0$) transition. A peak-fit analysis suggested that this bump consisted of several $1/2^-$ states.
We propose that this bump is due to the mirror state of the 13.5 MeV-state in $^{13}$N, which dominantly decays to the $\alpha$ condensed state in $^{12}$C. It was speculated that the $1/2^-$ states around $E_x = 12.5$ MeV were candidates for the $\alpha$ condensed state, but the $3\alpha + n$ orthogonality condition model suggests that the $\alpha$ condensed state is unlikely to emerge as the negative parity states.
We also found two $1/2^+$ or $3/2^+$ states at $E_x = 14.5$ and 16.1 MeV excited with the isoscalar dipole ($IS1$) strengths.
We suggest that the 16.1-MeV state is a possible candidate for the $\alpha$ condensed state predicted by the cluster-model calculations on the basis of the good correspondence between the experimental and calculated level structures.
However, the theoretical $IS1$ transition strength for this state is significantly smaller than the measured value. 
Further experimental information is strongly desired to establish the $\alpha$ condensed state in $^{13}$C.
\end{abstract}

\subjectindex{xxxx, xxx}

\maketitle

\section{Introduction}
The $\alpha$-cluster structures are well known to appear in atomic nuclei.
The cluster structures develop in nuclei when their energies are close to the cluster-decay thresholds.
This threshold rule is based on the fact that the interaction between $\alpha$ clusters is relatively weak, while protons and neutrons are tightly bound in the $\alpha$ clusters.
The Ikeda diagram schematically demonstrates this threshold rule in light self-conjugated $A=4k$ nuclei~\cite{Horiuchi1968}.
The most famous example of the $\alpha$-cluster state is the $0_2^+$ state at $E_x = 7.65$~MeV in $^{12}$C, known as the Hoyle state, which locates about 400 keV above the $3\alpha$-decay threshold.
The $3\alpha$-cluster model calculations reasonably well reproduce the energy of this $0_2^+$ state~\cite{Fujiwara1966, Uegaki1977, Kamimura1981}, although the shell-model calculation does not~\cite{Navratil2003}.
Therefore, the $0_2^+$ state is considered to have a spatially developed structure consisting of three $\alpha$ clusters unlike the ground state with a shell-model-like compact structure.

In the early days, the $0_2^+$ state was suggested to have a linear configuration of three $\alpha$ clusters~\cite{Morinaga1956, Morinaga1966}.
In recent years, it has been pointed out that this state is the $\alpha$ condensed state in which the three $\alpha$ clusters occupy the same $0S$ orbit like the atomic Bose-Einstein condensation~\cite{Tohsaki2001, Funaki2003, Yamada2005}.
According to the $3\alpha$ orthogonality condition model~(OCM) calculation~\cite{Yamada2005}, the three $\alpha$ clusters in the $0_2^+$ state are condensed into the $0S$ orbit with a high occupancy about 70\%.
This calculation suggests the momentum distribution of the $\alpha$ clusters has a sharp peak like a $\delta$-function around 0~fm$^{-1}$.
Reflecting this momentum distribution, the $0_2^+$ state is considered to be a low-density dilute cluster gas state with a spatially expanded density distribution.
The radius of the $0_2^+$ state is estimated to be about twice as large as the radius of the ground state.
The theoretical calculation with the Gross-Pitaevskii and Hill-Wheeler equations by Yamada and Schuck~\cite{Yamada2004} predicts that the $k\alpha$ condensed states with similar properties to the $0_2^+$ state can exist in heavier self-conjugated $A = 4k$ nuclei up to $^{40}$Ca.
The experimental searches for such $\alpha$ condensed states are being intensively carried out in $^{16}$O~\cite{Wakasa2007, Itoh2010, Ogloblin2016, Li2017}, $^{20}$Ne~\cite{Adachi2021}, $^{24}$Mg~\cite{Kawabata2013}, and $^{28}$Si~\cite{Bishop2019}.

It is of great interest whether the $\alpha$ condensed states exist in $A \neq 4k$ nuclei.
Kawabata {\it et al.}~\cite{Kawabata2007} have reported that the $3/2_3^-$ state at $E_x = 8.56$~MeV in $^{11}$B, locating about 100 keV below the $\alpha$-decay threshold, is excited from the ground state with the strong isoscalar monopole ($IS0$) strength.
The experimental value of the $IS0$ transition strength is $B(IS0,~\mathrm{g.s.} \rightarrow 3/2_3^-) = 96 \pm 16$~(fm$^4$)~\cite{Kawabata2007}, which is comparable to that from the ground state to the $0_2^+$ state in $^{12}$C, $B(IS0,~\mathrm{g.s.} \rightarrow 0_2^+) = 121\pm 9$~(fm$^4$)~\cite{Ajzenbergselove1991}.
This large $IS0$ transition strength is successfully explained by the antisymmetrized molecular dynamics~(AMD) calculation although the shell-model calculation cannot reproduce such a large $IS0$ transition. 
Therefore, the $3/2_3^-$ state in $^{11}$B has been proposed to have a spatially developed $2\alpha + t$ cluster structure akin to the $0_2^+$ state in $^{12}$C~\cite{Kawabata2007}.
This fact implies that the dilute nature of the $0_2^+$ state is preserved even if one $\alpha$ cluster is replaced by a triton.
Actually, the various nuclear model calculations have pointed out the dilute nature of the $3/2_3^-$ state~\cite{Yamada2010, Kanada-EnYo2015, Zhou2018}.
It was speculated that this state was the $\alpha$ condensed state with an additional $P$-wave triton. 
However, the theoretical calculations showed this state could not be regarded as the $\alpha$ condensed state with a triton since the $2\alpha$ clusters did not fully condense into the same $0S$ orbit.
Instead, the $\alpha$ condensed state in which the constituent clusters were condensed into the $0S$ orbit was predicted to be the $1/2^+$ state locating at $E_x = 11.85$ MeV about 750 keV above the $2\alpha + t$ threshold~\cite{Yamada2010}.
This predicted $1/2^+$ state has never been experimentally established, and the existence of the $\alpha$ condensed state in $^{11}$B is still under discussion.

It is interesting to see whether the $\alpha$ condensed nature of the $0_2^+$ state in $^{12}$C still remains when an extra neutron is added to the $0_2^+$ state.
The full four-body $3\alpha + n$ OCM calculation predicts that there exist several $1/2^{\pm}$ excited states in $^{13}$C with spatially developed cluster structures near the $3\alpha + n$ threshold at $E_x = 12.2$ MeV~\cite{Yamada2015}.
An excess $P$-wave neutron couples to the $^{12}$C($0_2^+$) core in the $1/2^-$ states, 
whereas an $S$-wave neutron couples to the core in the $1/2^+$ states.
This calculation proposes that the $1/2_5^+$ state predicted at $E_x = 14.9$~MeV is the $\alpha$ condensed state with an excess neutron where the three $\alpha$ clusters and the excess neutron occupy the $(0S)^3_{\alpha}(S)_\nu$ configuration.
Similar results are obtained by the AMD calculation~\cite{Chiba2020}, in which the $1/2_2^+$ state predicted at $E_x = 15.4$ MeV is proposed as the $\alpha$ condensed state.
However, the existence of these $1/2^{\pm}$ states has not been experimentally confirmed yet.
Since the spin and parity of the ground state in $^{13}$C are $1/2^-$, the $1/2^-$ and $1/2^+$ states are excited through the $IS0$ and $IS1$ transitions, respectively.
Therefore, it is important to search for the $1/2^{\pm}$ states slightly above the $3\alpha + n$ threshold and compare the measured $IS0$ and $IS1$ transition strengths with the theoretical predictions in order to examine whether the $\alpha$ condensed state is still preserved in $^{13}$C, irrespective of additional one neutron.

The $\alpha$ inelastic scattering is a powerful tool to extract isoscalar transition strengths with various transferred angular momenta ($\Delta L$)~\cite{Adachi2018, PhysRevC.33.1116, Youngblood1999, Itoh2002, Uchida2003, Itoh2003, Uchida2004, Li2007, Wakasa2007, Itoh2011, Itoh2013, Gupta2016, Peach2016, GiantResonance}.
Since the spin and isospin of $\alpha$ particles are zero, the natural-parity isoscalar transitions are selectively induced in the $\alpha$ inelastic scattering if the Coulomb interaction is neglected.
The reaction process in the $\alpha$ inelastic scattering is well described by the distorted-wave Born-approximation (DWBA) calculation with the single-folding potential.
Thus, the cross sections are approximately proportional to the transition strengths.
Since the $\alpha$ inelastic scattering is a surface reaction, 
the angular distribution of the cross section is not sensitive to the details of the internal wave function and is characterized by the transferred angular momentum.
Therefore, the multipole decomposition analysis~(MDA) works well to extract the different $\Delta L$ components from the measured cross sections~\cite{Itoh2002,Uchida2003,Itoh2003,Uchida2004,Li2007,Itoh2011,Itoh2013,Gupta2016,Peach2016}.
Because the angular distributions of the cross sections are quite different near $\theta_{\mathrm{c.m.}} = 0^\circ$ depending on the transferred angular momenta, it is desirable to measure the $\alpha$ inelastic scattering at forward angles including 0 degrees in order to conduct the MDA.

In the present work, we studied the nuclear structures of $^{13}$C via the $^{13}\mathrm{C}(\alpha,\alpha')$ reaction at $E_\alpha = 388$ MeV at forward angles including 0 degrees.
We compared the measured cross sections with the DWBA calculations with the single-folding potentials to determine the isoscalar transition strengths in $^{13}$C.
We extracted the strength distributions for the isoscalar $\Delta L = 0$--3 transitions by the MDA.
We compared the experimental level diagram and transition strengths for the $1/2^\pm$ states excited through the $IS0$ and $IS1$ transitions with theoretical predictions.
We discussed the structures of these states and the existence of the $\alpha$ condensed state in $^{13}$C.

\section{Experiment}
\begin{figure*}[b]
 \begin{center}
  \includegraphics[width=\linewidth]{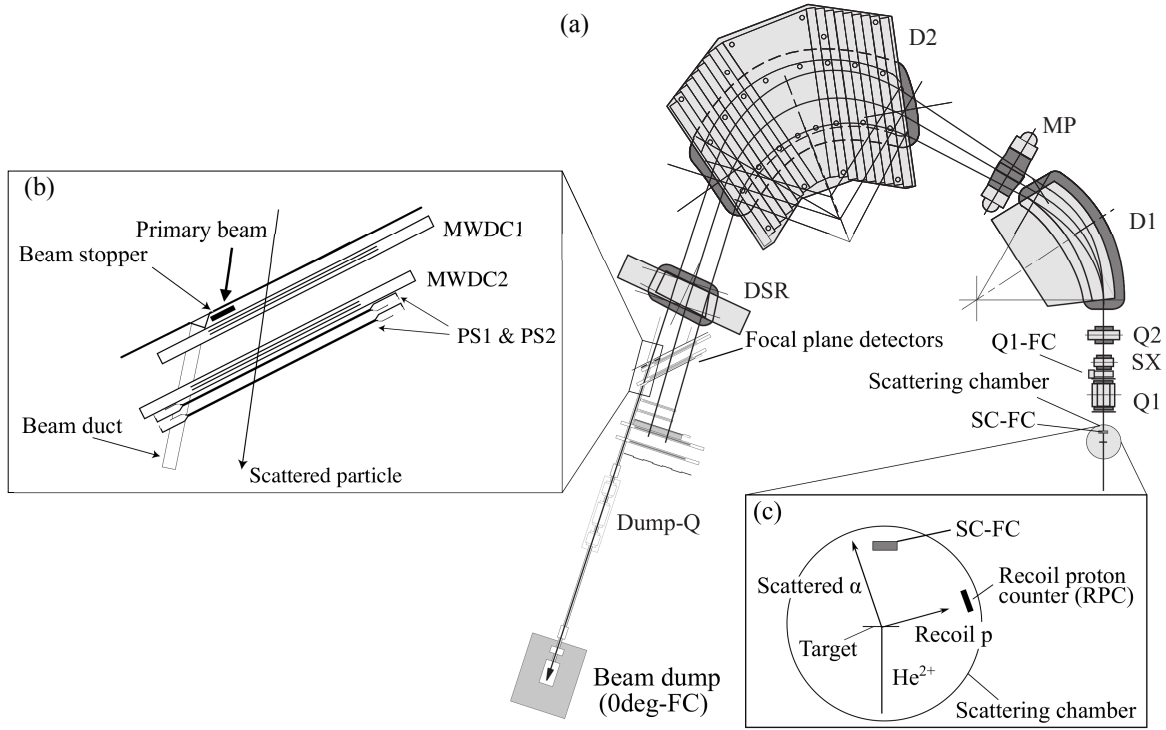}
  \caption{(a) Overview of the experimental setup at the scattering angle of $\theta_{\mathrm{lab}} = 0^\circ$.
(b) Expanded view of the focal plane detectors to measure the low-lying states at $\theta_{\mathrm{lab}} = 0^\circ$. A beam stopper is installed upstream of the focal plane detectors.
(c) Expanded view of the detector layout in the scattering chamber of the Grand Raiden spectrometer.}
  \label{setup}  
 \end{center}
\end{figure*}

We performed the experiment at Research Center for Nuclear Physics~(RCNP) in Osaka University.
An incident He$^{2+}$ beam extracted from the Neomafious ion source was accelerated by the AVF and ring cyclotrons up to $E_\alpha = $ 388 MeV, and was transported to the west experimental hall through the WS beam line~\cite{Wakasa2002}.
The beam was achromatically focused on a self-supporting 98\% enriched $^{13}$C target with a thickness of 1.5~mg/cm$^2$ in the scattering chamber of the high-resolution magnetic spectrometer Grand Raiden~\cite{Fujiwara1999}.
The size of the beam spot was about 1~mm in diameter on the target.
The experimental setup of the Grand Raiden spectrometer for the measurement at $\theta_{\mathrm{lab}} = 0^\circ$ is shown in Fig.~\ref{setup}(a).
Scattered alpha particles were momentum analyzed with the Grand Raiden spectrometer and detected by the focal plane detectors, which consisted of the two multi-wire drift chambers~(MWDCs) and the two plastic scintillation counters (PS1 and PS2).
Scattering angles were reconstructed from the particle trajectories on the focal plane.
Trigger signals were generated by coincidence signals of the PS1 and PS2 installed downstream of the MWDCs.
Particle identification was performed using the correlations between the momentum and time-of-flight and between the momentum and energy deposition to PS1 by scattered particles.
Cross sections for the $\alpha$ elastic and inelastic scattering from the $^{13}$C target were measured at $\theta_{\mathrm{lab}} = 0$--$20.2^\circ$.

\begin{figure}[b]
\begin{center}
\includegraphics[width=0.8\linewidth]{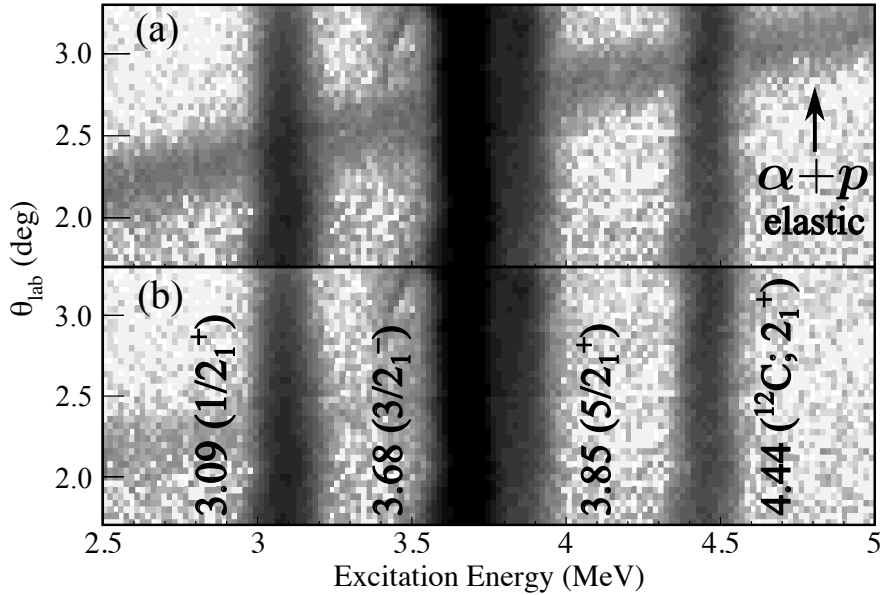}
\caption{Two-dimensional scatter plot of the scattering angle in the laboratory frame versus the excitation energy of $^{13}$C measured (a) without and (b) with the recoil proton counter. The four vertical loci correspond to the states at $E_x = 3.09,~3.68$, and 3.85~MeV in $^{13}$C and the state at $E_x = 4.44$~MeV in $^{12}$C. The locus indicated by the arrow corresponds to the $\alpha + p$ elastic scattering.}
\label{recoil_proton}
\end{center}  
\end{figure}

In the measurement at $\theta_{\mathrm{lab}} = 0^\circ$, the He$^{2+}$ beam was guided to the focal plane through the Grand Raiden spectrometer in the same way as scattered alpha particles.
The beam current was measured by a Faraday cup (0deg-FC) in the beam dump as shown in Fig.~\ref{setup}(a). 
The excitation-energy region measurable with this setup was limited to $E_x \ge 7$~MeV due to the geometrical relationship between the beam duct and the sensitive area of the focal plane detectors.
Therefore, we prepared another experimental setup to extend the measurable excitation-energy region down to $E_x = 2.5$~MeV.
As shown in Fig.~\ref{setup}(b), an alternative beam stopper was installed upstream of the MWDCs to make the beam position as close as possible to the sensitive area of the focal plane detectors.
Since this beam stopper could not measure the beam current, the absolute value of the cross sections was not determined.
Therefore, the cross section measured in this setup was normalized using the $1/2_2^-$ state at $E_x = 8.86$~MeV because this state could be measured in both of the two setups at $\theta_{\mathrm{lab}} = 0^\circ$.
Different Faraday cups behind the Q1 magnet (Q1-FC) and in the scattering chamber (SC-FC) were used for the measurements at the backward angles of $\theta_{\mathrm{lab}} = 1.7$--5.9$^\circ$ and $\theta_{\mathrm{lab}} = 6.0$--20.2$^\circ$, respectively.
These Faraday cups were remotely moved to avoid intercepting the beam and scattered alpha particles when they were not used.

Figure \ref{recoil_proton} shows a two-dimensional scatter plot of the scattering angle versus the excitation energy of $^{13}$C measured at $\theta_{\mathrm{lab}} = 1.7$--$3.2^\circ$.
The $\alpha + p$ elastic scattering events due to hydrogen contaminants in the $^{13}$C target are overlapped with inelastic events due to the discrete states in $^{13}$C and $^{12}$C as seen in Fig.~\ref{recoil_proton}(a).
We used a recoil proton counter~(RPC) to detect recoil protons and to tag the background events caused by the $\alpha$ + $p$ elastic scattering.
The RPC was an $E$--$\Delta E$ telescope consisting of two plastic scintillators. 
The sizes of both scintillators were 35 mm wide and 240 mm high, and the thicknesses were 1 mm and 9 mm, respectively.
Particle identification of recoil particles was performed using the $E$--$\Delta E$ correlation.
Figure \ref{setup}(c) shows the expanded view of the experimental setup in the scattering chamber.
The RPC was installed in the scattering chamber, and its installation angle was remotely changed to catch recoil protons depending on the setting angle of the Grand Raiden spectrometer.
The background from the hydrogen contaminants in the target was rejected by identifying the events in which the RPC detected recoil protons as shown in Fig.~\ref{recoil_proton}(b).

We also performed a measurement using a $^{\mathrm{nat}}$C target with the same setup for $^{13}$C to subtract background contributions due to the $^{12}$C impurities in the $^{13}$C target.
Two $^{\mathrm{nat}}$C targets with different thicknesses of 0.5~mg/cm$^2$ and 2.8~mg/cm$^2$ were used at $\theta_{\mathrm{lab}} = 0$--$5.9^\circ$ and $\theta_{\mathrm{lab}} = 6.0$--$20.2^\circ$, respectively.
Since the background measurement with the $^{\mathrm{nat}}$C target was not performed for the low-lying states in the excitation-energy region of $E_x \le 7$ MeV at $\theta_{\mathrm{lab}} = 0^\circ$, 
the $^{12}$C contribution to the measurement for the $^{13}\mathrm{C}(\alpha, \alpha')$ reaction was evaluated only at $E_x \ge 7$~MeV.
Therefore, the background due to the first excited state at $E_x = 4.44$~MeV in $^{12}$C was not subtracted at $\theta_{\mathrm{lab}} = 0^\circ$.
However, its contribution was clearly distinguished from the discrete states of $^{13}$C in the excitation-energy spectrum.
The cross sections for the $\alpha$ elastic scattering off $^{12}$C were also measured.

\begin{figure}[b]
\begin{center}
\includegraphics[width=0.65\linewidth]{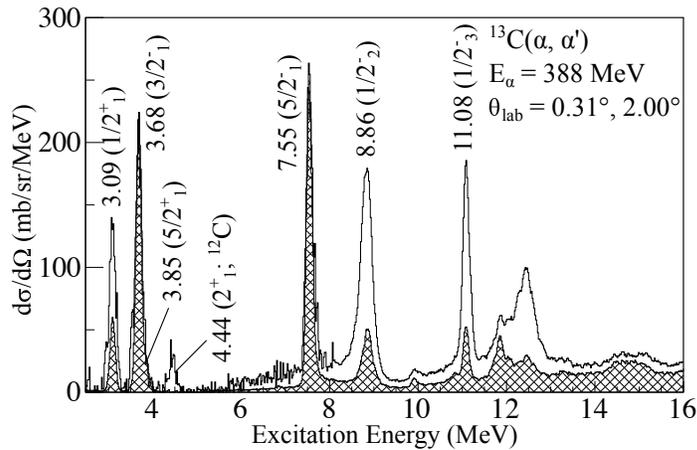}
\caption{Typical excitation-energy spectra for the $^{13}$C($\alpha, \alpha'$) reaction measured at $\theta_{\mathrm{lab}} = 0.31^\circ$ and $2.00^\circ$. The open and hatched histograms show the spectra at $\theta_{\mathrm{lab}} = 0.31^\circ$ and $2.00^\circ$, respectively.}
\label{single_spectrum}
\end{center}  
\end{figure}

\section{Analysis and results}
\subsection{Excitation-energy spectra}
Typical excitation-energy spectra for the $^{13}$C($\alpha, \alpha'$) reaction measured at $\theta_{\mathrm{lab}} = 0.31^\circ$ and $2.00^\circ$ are shown in Fig.~\ref{single_spectrum}.
The open and hatched histograms show the spectra at $\theta_{\mathrm{lab}} = 0.31^\circ$ and $2.00^\circ$, respectively.
Several discrete states in the excitation-energy region of $E_x \le 11.08$~MeV listed in the compilation~\cite{Ajzenbergselove1991} are clearly observed in the spectra.
The energy resolution for the first excited state at $E_x = 3.09$~MeV was about 140~keV at full width at half maximum, dominated by the energy spread of the He$^{2+}$ beam.

\begin{figure}[t]
 \begin{center}
\includegraphics[width=\linewidth]{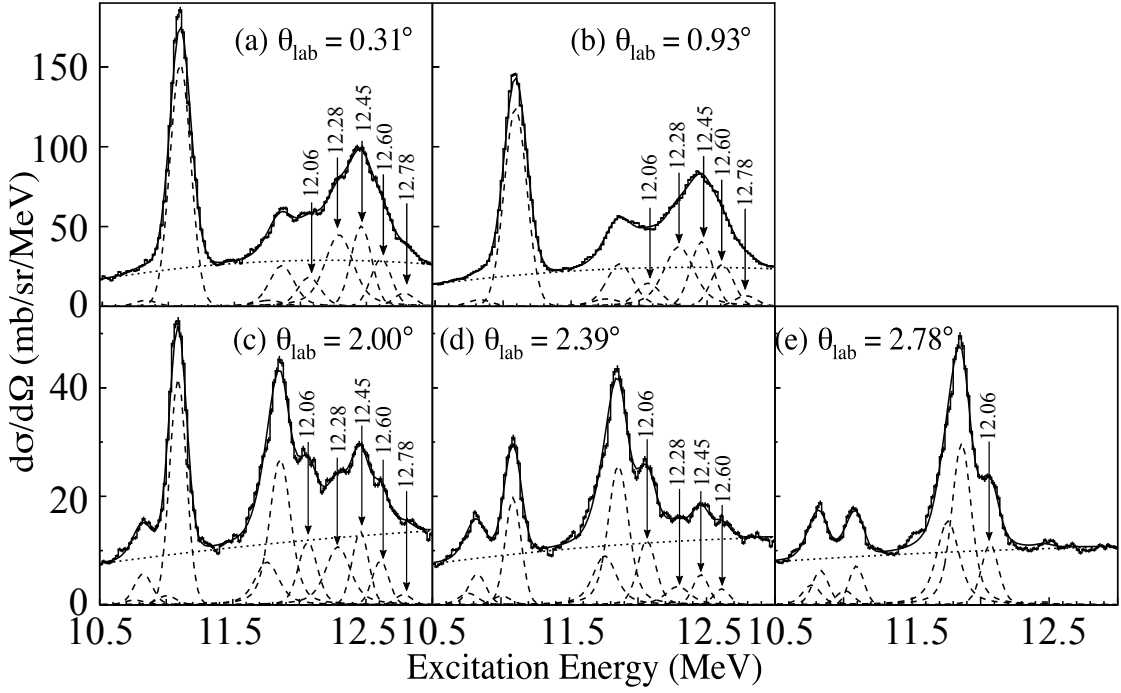}
\caption{Excitation-energy spectra in the region of  $10.5~\mathrm{MeV} \le E_x \le 13.0~\mathrm{MeV}$ measured at $\theta_{\mathrm{lab}} = 0.31$--2.78$^\circ$. Only statistical errors were taken into account in the peak-fit analysis. The solid lines show the result of the fit. The dashed lines show the fitting curves for the excited states. The dotted lines show the assumed continuum states. The new states included in the peak-fit analysis are indicated by the vertical arrows.}
\label{spectrum_fitting} 
\end{center} 
\end{figure}

In addition to the discrete states at $E_x \le 11.08$~MeV, we observed a prominent bump structure around $E_x = 12.5$~MeV in the spectrum at $\theta_{\mathrm{lab}} = 0.31^\circ$.
The detailed level structures at $10.5~\mathrm{MeV} \le E_x \le 13.0$ MeV were examined by fitting the measured spectra with the known states in Ref.~\cite{Ajzenbergselove1991} and several new states.
The shape of the excited states was assumed to be the Voigtian shape in which the Lorentzian profile was convoluted with the Gaussian profile as
\begin{equation}
 V(E\hbox{;}~\sigma, \Gamma) = \int_{-\infty}^{\infty} G(E'\hbox{;}~\sigma)L(E-E'\hbox{;}~\Gamma)dE'.
\end{equation}
Here, $G$ is the Gaussian profile describing the beam shape and $L$ is the Lorentzian profile.
$E$ and $\Gamma$ denote the energy and width of the excited state. $\sigma$ denotes the energy spread of the beam.
A quadratic polynomial was used to describe continuum states.
The energies and widths of the known states were fixed at the values listed in Ref.~\cite{Ajzenbergselove1991}, whereas those of the new states were determined to fit the five spectra obtained at $\theta_{\mathrm{lab}} = 0.31$--$2.78^\circ$ simultaneously.
Five new states were introduced to reasonably well fit the measured spectra.
Although we included all the states listed in Ref.~\cite{Ajzenbergselove1991} in the peak-fit analysis as tabulated in the left three columns in Table \ref{Tab:spectra_12_5}, several states with almost zero yields were neglected in the following analysis.  
The fit results are drawn by the solid lines in Fig.~\ref{spectrum_fitting}. 
The energies and widths of the excited states included in the peak-fit analysis are summarized in the fourth and fifths columns of Table \ref{Tab:spectra_12_5}.
The isoscalar transition strengths $B(IS\lambda)$ and the energy-weighted sum-rule (EWSR) fractions for these states determined by the DWBA analysis are also given in Table \ref{Tab:spectra_12_5}.
The details of the DWBA analysis will be described in Sec.~\ref{sec:dwba}.

\begin{landscape}
\begin{table}[t]
\begin{center}
\begin{threeparttable}
 \caption{Excitation energies and widths of the excited states at 10.5~MeV $\le E_x \le $ 13.0 MeV taken from Ref~\cite{Ajzenbergselove1991} and those included in the peak-fit analysis. Isoscalar transition strengths $B(IS\lambda)$ and EWSR fractions determined by the DWBA analysis are also tabulated.} 
 \label{Tab:spectra_12_5}
 \begin{tabularx}{\linewidth}{CCCCCCCCC}\hline\hline
\multicolumn{3}{c}{Ref.~\cite{Ajzenbergselove1991}} & & \multicolumn{5}{c}{Present}\\ \cline{1-3} \cline{5-9}
\multicolumn{1}{c}{$J^{\pi}$}& \multicolumn{1}{c}{$E_x$}&\multicolumn{1}{c}{$\Gamma$} & &\multicolumn{1}{c}{$E_x$} & \multicolumn{1}{c}{$\Gamma$}&\multicolumn{1}{c}{Multipolarity}&\multicolumn{1}{c}{$B(IS\lambda)$} & \multicolumn{1}{c}{EWSR fraction}\\ 
 & \multicolumn{1}{c}{(MeV)} & \multicolumn{1}{c}{(keV)} & & \multicolumn{1}{c}{(MeV)} & \multicolumn{1}{c}{(keV)} & & \multicolumn{1}{c}{(fm$^{2\lambda}$)\tnote{a}}  & \multicolumn{1}{c}{(\%)}\\ \hline

\multicolumn{1}{c}{$7/2^-$} & \multicolumn{1}{c}{$10.753\pm 0.004$} & \multicolumn{1}{c}{$55\pm 2$} & & \multicolumn{1}{c}{10.753} & \multicolumn{1}{c}{55} & \multicolumn{1}{c}{$IS$4} & \multicolumn{1}{c}{$370\pm 140$} & \multicolumn{1}{c}{$0.08 \pm 0.03$}\\  

\multicolumn{1}{c}{$5/2^-$} & \multicolumn{1}{c}{$10.818\pm 0.005$} & \multicolumn{1}{c}{$24\pm 3$} & & \multicolumn{1}{c}{10.818} & \multicolumn{1}{c}{24} & \multicolumn{1}{c}{$IS$2} & \multicolumn{1}{c}{$1.7\pm 0.1$} & \multicolumn{1}{c}{$0.27 \pm 0.02$}\\

\multicolumn{1}{c}{$1/2^+$} & \multicolumn{1}{c}{$10.996\pm 0.006$} & \multicolumn{1}{c}{$37\pm 4$} & & \multicolumn{1}{c}{10.996} & \multicolumn{1}{c}{37} & \multicolumn{1}{c}{$IS$1} & \multicolumn{1}{c}{$< 2.3$} & \multicolumn{1}{c}{$ < 0.5$}\\

\multicolumn{1}{c}{$1/2^-$} & \multicolumn{1}{c}{$11.080\pm 0.005$} & \multicolumn{1}{c}{$< 4$} & & \multicolumn{1}{c}{11.080} & \multicolumn{1}{c}{$4$} & \multicolumn{1}{c}{$IS$0} & \multicolumn{1}{c}{$19.2\pm 0.3$} & \multicolumn{1}{c}{$3.2\pm 0.1$}\\ 

\multicolumn{1}{c}{$3/2^-$} & \multicolumn{1}{c}{$11.748\pm 0.010$} & \multicolumn{1}{c}{$110\pm 15$} & & \multicolumn{1}{c}{11.748} & \multicolumn{1}{c}{110} & \multicolumn{1}{c}{$IS$2} & \multicolumn{1}{c}{$5.6\pm 1.6$}& \multicolumn{1}{c}{$1.0 \pm 0.3$} \\

\multicolumn{1}{c}{$7/2^+$} & \multicolumn{1}{c}{$11.848\pm 0.004$} & \multicolumn{1}{c}{$68\pm 4$} & & \multicolumn{1}{c}{11.848} & \multicolumn{1}{c}{68} & \multicolumn{1}{c}{$IS$3} & \multicolumn{1}{c}{$740\pm 80$} & \multicolumn{1}{c}{$4.8 \pm 0.5$} \\  

\multicolumn{1}{c}{$5/2^+$} & \multicolumn{1}{c}{$11.950\pm 0.040$} & \multicolumn{1}{c}{$500\pm 80$} & & \multicolumn{1}{c}{} & \multicolumn{1}{c}{} & \multicolumn{1}{c}{} & \multicolumn{1}{c}{} & \multicolumn{1}{c}{} \\  

\multicolumn{1}{c}{} & \multicolumn{1}{c}{} & \multicolumn{1}{c}{} & & \raisebox{0.5em}{$12.055 \pm 0.001$} & \raisebox{0.5em}{$38\pm 4$} & \shortstack{$IS$0\tnote{b} \\ $IS2$\tnote{b}} & \shortstack{$1.7\pm 0.3$ \\ $2.2\pm 0.1$ } & \shortstack{$0.32\pm 0.06$ \\ $0.40\pm 0.02$} \rule[0mm]{0mm}{10mm}\\ 

\multicolumn{1}{c}{$3/2^+$} & \multicolumn{1}{c}{$12.106\pm 0.005$} & \multicolumn{1}{c}{$540\pm 70$} & & \multicolumn{1}{c}{} & \multicolumn{1}{c}{} & \multicolumn{1}{c}{} & \multicolumn{1}{c}{} & \multicolumn{1}{c}{} \\

\multicolumn{1}{c}{$5/2^-$} & \multicolumn{1}{c}{$12.130\pm 0.050$} & \multicolumn{1}{c}{$80\pm 30$} & & \multicolumn{1}{c}{} & \multicolumn{1}{c}{} & \multicolumn{1}{c}{} & \multicolumn{1}{c}{} & \multicolumn{1}{c}{} \\

\multicolumn{1}{c}{$1/2^+$} & \multicolumn{1}{c}{$12.140\pm 0.070$} & \multicolumn{1}{c}{$430\pm 70$} & & \multicolumn{1}{c}{} & \multicolumn{1}{c}{} & \multicolumn{1}{c}{$IS$1} & \multicolumn{1}{c}{$< 1.8$\tnote{c}} & \multicolumn{1}{c}{$< 0.5$\tnote{c}} \\ 

\multicolumn{1}{c}{$3/2^-$} & \multicolumn{1}{c}{$12.187\pm 0.010$} & \multicolumn{1}{c}{$150\pm 40$} & & \multicolumn{1}{c}{} & \multicolumn{1}{c}{} & \multicolumn{1}{c}{}& \multicolumn{1}{c}{} & \multicolumn{1}{c}{} \\  

\multicolumn{1}{c}{} & \multicolumn{1}{c}{} & \multicolumn{1}{c}{} & & \multicolumn{1}{c}{$12.282 \pm 0.005$} & \multicolumn{1}{c}{$122\pm 22$} & \multicolumn{1}{c}{$IS$0}&\multicolumn{1}{c}{$4.1\pm 0.4$} &\multicolumn{1}{c}{$0.77\pm 0.08$} \\ 

\multicolumn{1}{c}{$7/2^-$} & \multicolumn{1}{c}{$12.438\pm 0.012$} & \multicolumn{1}{c}{$140\pm 30$} & & \multicolumn{1}{c}{} & \multicolumn{1}{c}{} & \multicolumn{1}{c}{} &\multicolumn{1}{c}{} &\multicolumn{1}{c}{}  \\ 

\multicolumn{1}{c}{} & \multicolumn{1}{c}{} & \multicolumn{1}{c}{} & & \multicolumn{1}{c}{$12.450\pm0.003$} & \multicolumn{1}{c}{$\ll 70$\tnote{d}} & \multicolumn{1}{c}{$IS$0} &\multicolumn{1}{c}{$4.9\pm 0.4$} &\multicolumn{1}{c}{$0.93\pm 0.08$} \\ 

\multicolumn{1}{c}{} & \multicolumn{1}{c}{} & \multicolumn{1}{c}{} & & \multicolumn{1}{c}{$12.601\pm0.003$} & \multicolumn{1}{c}{$\ll 70$\tnote{d}} & \multicolumn{1}{c}{$IS$0} &\multicolumn{1}{c}{$3.1\pm 0.2$} &\multicolumn{1}{c}{$0.59\pm 0.03$} \\ 

\multicolumn{1}{c}{} & \multicolumn{1}{c}{} & \multicolumn{1}{c}{} & & \multicolumn{1}{c}{$12.775\pm 0.004$} & \multicolumn{1}{c}{$\ll 70$\tnote{d}} & \multicolumn{1}{c}{$IS$0} &\multicolumn{1}{c}{$0.92\pm 0.05$} &\multicolumn{1}{c}{$0.18\pm 0.01$} \\ \hline \hline
 \end{tabularx}
\begin{tablenotes}
 \item[a] The units of $B(IS0)$ and $B(IS1)$ are fm$^4$ and fm$^6$, respectively.
 \item[b] Both the $IS0$ and $IS2$ transitions contributed to this state. See text for detail.
 \item[c] Although this state was not included in the peak-fit analysis, the upper limits of the 95\% confidence intervals for the $B(IS1)$ and EWSR fraction were evaluated. See text for detail.
 \item[d] Smaller enough than the energy resolution ($\sigma \sim 70$ keV) in the present measurement.
\end{tablenotes}
\end{threeparttable}
\end{center}
\end{table} 
\end{landscape}

\subsection{DWBA calculation} \label{sec:dwba}
We calculated cross sections for the $^{13}\mathrm{C}(\alpha, \alpha')$ reaction at $E_\alpha = 388$ MeV in the framework of the DWBA using a computer code ECIS-95~\cite{ECIS95}.
The calculated cross sections were compared with the measured cross sections to determine the isoscalar transition strengths in $^{13}$C.
The optical-model potential for the $\alpha$ elastic scattering was used as the distorting potential.
The same distorting potential was applied to the entrance and exit channels.
The optical-model potential was obtained using the single-folding model, where an effective $\alpha N$ interaction was folded with the ground-state density distribution of the target nucleus.

The proton-density distributions $\rho_p(r)$ for the ground state of $^{12}$C and $^{13}$C were derived from the electron-scattering data~\cite{DeVries1987}.
The neutron-density distribution $\rho_n(r)$ of $^{12}$C was assumed as the same shape as the proton-density distribution of $^{12}$C.
The neutron-density distribution of $^{13}$C was assumed to be $ \rho_n(r) = \rho_p(r')$ with $r' = (6/7)^{1/3}r$. 
Many previous analyses of the $\alpha$ inelastic scattering~\cite{Youngblood1999, Itoh2002, Uchida2003, Itoh2003, Uchida2004, Wakasa2007, Kawabata2007, Li2007, Itoh2011, Itoh2013, Gupta2016, Peach2016} employed the density-dependent effective $\alpha N$ interaction proposed by Satchler and Khoa~\cite{Satchler1997}.
However, the recent systematic analysis of the $\alpha$ inelastic scattering demonstrated that the cross sections for the $IS0$ transitions tended to be overestimated when the density-dependent interaction was used~\cite{Adachi2018}.
A more recent analysis of the $\alpha$ inelastic scattering off $^{10}$C employed the density-independent interaction to deduce the $IS2$ transition strength from the ground state to the $2_1^+$ state~\cite{Furuno2019}.
Therefore, we used the density-independent interaction parameterized by the Gaussian functions as 
\begin{equation}
 u(|\bm{r}-\bm{r'}|) = -ve^{-|\bm{r}-\bm{r'}|^2/\alpha^2_v}-iwe^{-|\bm{r}-\bm{r'}|^2/\alpha^2_w}
\end{equation}
in the present DWBA analysis.
Here, $v$ and $w$ are the depth parameters.
$\alpha_v$ and $\alpha_w$ are the range parameters of the interaction.

\begin{figure}[t]
\begin{center}
\includegraphics[width=\linewidth]{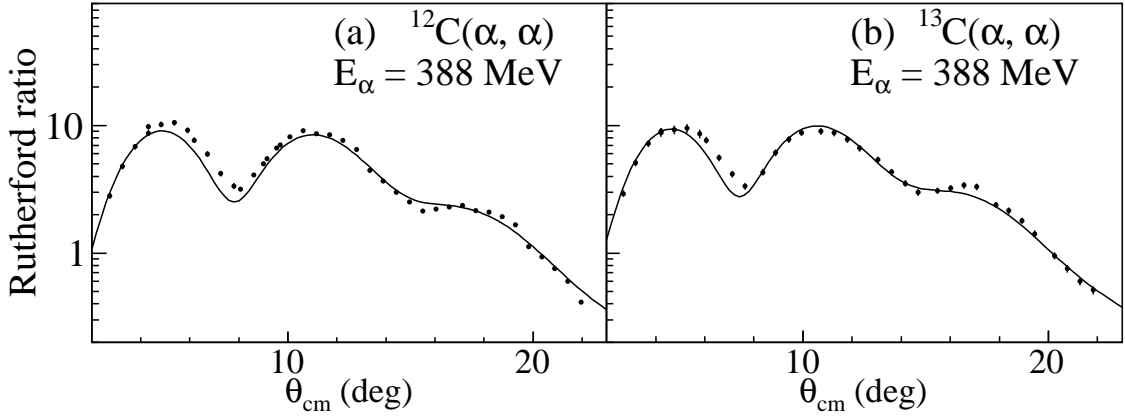}
\caption{Angular distributions of the cross sections for the $\alpha$ elastic scattering off (a) $^{12}$C and (b) $^{13}$C at $E_\alpha = 388$~MeV relative to the Rutherford cross sections~(Rutherford ratio). The measured cross sections (the solid circles) are compared with the calculations (the solid lines) using the single-folding potentials.}
\label{elastic}
\end{center}
\end{figure}

The interaction parameters were determined to reproduce the measured cross section of the $\alpha$ elastic scattering off $^{12}$C as shown in Fig.~\ref{elastic}(a) since the ambiguity in the assumption about the neutron-density distribution of $^{13}$C caused uncertainty in the determination of the interaction.
The depth and range parameters were determined to be $v = 20.4$~MeV, $w = 9.35$~MeV, $\alpha_v = 1.97$~fm, and $\alpha_w = 2.18$~fm.
Figure~\ref{elastic}(b) shows the comparison between the measured elastic cross section for $^{13}$C and the calculated cross section using the same interaction parameters with $^{12}$C.
The measured cross section for $^{13}$C is well reproduced by the single-folding potential, indicating that the assumption on the neutron-density distribution of $^{13}$C and application of the same effective interaction for $^{12}$C and $^{13}$C are reasonable.

The transition potentials for the $\alpha$ inelastic scattering were calculated by folding the effective $\alpha N$ interaction with the transition densities.
The transition densities $\delta \rho^{(\lambda)}(r)$ for the $\Delta L = \lambda$ transitions were obtained by the macroscopic model~\cite{Harakeh1981, Satchler1987} as  
\begin{align}
 \delta\rho^{(0)}(r) & =  -\alpha_0\left(3+r\frac{d}{dr}\right)\rho(r)~~(\lambda = 0),\label{ts_0} \\ 
 \delta\rho^{(1)}(r) & =  -\frac{\beta_1}{\sqrt{3}R}\left[3r^2\frac{d}{dr} + 10r- \frac{5}{3}\langle r^2\rangle\frac{d}{dr} + \epsilon\left(r\frac{d^2}{dr^2} + 4\frac{d}{dr}\right)\right]\rho(r)~~(\lambda = 1),\label{ts_1} \\
 \delta\rho^{(\lambda)}(r) & = -\delta_\lambda\frac{d}{dr}\rho(r)~~(\lambda \ge 2),\label{ts_2}
\end{align}
 where $\epsilon = (4/E_{\mathrm{ISGQR}} + 5/E_{\mathrm{ISGMR}})\hbar^2/3mA$.
$E_{\mathrm{ISGQR}}$ and $E_{\mathrm{ISGMR}}$ are the mean excitation energies of the isoscalar giant quadrupole and monopole resonances. $\alpha_0$ is the deformation parameter, and $\delta_\lambda$ is the deformation length.
$\beta_1$ and $R$ are the collective coupling parameter for the isoscalar dipole resonance and the half-density radius of the Fermi mass distribution, respectively.
$\rho(r)$ is the density distribution of the ground state given by $\rho_n(r) + \rho_p(r)$.
The isoscalar transition matrix elements $M(IS\lambda)$ were calculated from the transition densities as 
\begin{align}
 M(IS0) & = \sqrt{4\pi}\int\delta\rho^{(0)}(r)r^4dr, \\
 M(IS1) & = \int\delta\rho^{(1)}(r)r^5dr, \\
 M(IS\lambda) & = \int\delta\rho^{(\lambda)}(r)r^{\lambda+2}dr, \qquad (\lambda \ge 2).
\end{align} 
The isoscalar transition strengths $B(IS\lambda)$ were obtained from $M(IS\lambda)$ using the relation as 
\begin{equation}
 B(IS\lambda) = \frac{1}{2J_i + 1}|M(IS\lambda)|^2,
\end{equation}
where $J_i$ is the spin of the initial state.
The transition strengths were deduced by determining $\alpha_0$, $\beta_1$, and $\delta_\lambda$ to reproduce the measured cross sections.
If one excited state at the excitation energy $E_x$ exhausts 100\% of the EWSR strength for the $\Delta L = \lambda$ transition, $\alpha_0$, $\beta_1$, and $\delta_\lambda$ are given by the sum-rule limits in Refs.~\cite{Harakeh1981, Satchler1987} as 
\begin{align}
 \alpha_0^2 &= \frac{2\pi\hbar^2}{mAE_x\langle r^2 \rangle}, \label{limit_0} \\ 
 \beta_1^2  &= \frac{6\pi\hbar^2}{mAE_x}\frac{R^2}{11\langle r^4 \rangle -25/3 {\langle r^2 \rangle}^2 -10\epsilon \langle r^2 \rangle}, \label{limit_1} \\ 
 \delta_\lambda^2 &= \frac{\lambda(2\lambda + 1)^2}{(\lambda + 2)}\frac{2\pi\hbar^2}{mAE_x}\frac{\langle r^{2\lambda - 2} \rangle}{{\langle r^{\lambda - 1} \rangle}^2}. \label{limit_2} 
\end{align}
The measured transition strengths were converted into the EWSR fractions using the sum-rule limits given by Eqs.~\eqref{limit_0}--\eqref{limit_2}.

Although the isoscalar transitions are selectively induced in the $\alpha$ inelastic scattering, the isovector transitions can be induced through the Coulomb interaction.
Especially, the isovector $E1$ transition is not negligible.
Therefore, the cross sections for the $\Delta L = 1$ transition were calculated by taking into account the interference between the isoscalar and isovector transitions. 
The isovector $E1$ strengths were determined from the photonuclear cross section or the radiative-decay width~\cite{Ajzenbergselove1991, Endt1993, Jury1979, Zubanov1983}.
Assuming that the $E1$ transition dominates the photonuclear cross section, the photonuclear cross section is related to the $E1$ strengths function as 
\begin{equation}
 \sigma (E) = \frac{16\pi^3}{9}\frac{E}{\hbar c}\frac{dB(E1)}{dE}. \label{csbe1}
\end{equation}

It should be commented that the validity of the DWBA analysis with the macroscopic model for the isoscalar $\Delta L = 0,~2$, and 3 transitions has been investigated by comparing the known electromagnetic transition strength $B(E\lambda)$ with the measured $B(IS\lambda)$ as in Refs.~\cite{PhysRevC.55.285, Adachi2018}.
However, for the isoscalar $\Delta L = 1$ transition, the validity has not been checked since the relation between $B(E1)$ and $B(IS1)$ is not straightforward.

\subsection{DWBA analysis for the discrete states}\label{sec:DWBA}
Figure~\ref{descrete_state} shows the comparison between the DWBA cross sections and the measured cross sections for the low-lying states at $E_x \le 8.86$ MeV as well as those for the high-lying states at $10.5~\mathrm{MeV}\le E_x \le 13.0$ MeV obtained by the peak-fit analysis. 
The measured cross sections for the low-lying states are reasonably well reproduced by the DWBA calculations for the corresponding $\Delta L = \lambda$ transition.
The steep increase of the cross section at forward angles for the 3.09-MeV state is fairly well reproduced by taking into account the isovector $E1$ transition by the Coulomb excitation as shown in Fig.~\ref{descrete_state}(a).
For the several states at $E_x \ge 10.5$ MeV whose energies are close to each other, we could not reliably separate these states in the peak-fit analysis. Therefore, the sum of the cross sections for these states are plotted as shown in Figs.~\ref{descrete_state}(f), (g), and (h).
These summed cross sections were fitted by the DWBA cross sections for the different $\Delta L$ transitions, and their transition strengths were separately determined.

The transition strengths and EWSR fractions determined by the present DWBA analysis for the high-lying states are listed in Table \ref{Tab:spectra_12_5}.
Those for the low-lying states are separately listed in Table \ref{strength}.
The systematic uncertainties in the present DWBA analysis were evaluated by a similar method to Ref.~\cite{Zenihiro2010}.
The 68\% confidence interval for the transition strength was determined by changing the amplitude of the transition density over the range satisfying 
\begin{equation}
 \tilde{\chi}^2 \le \tilde{\chi}_{\mathrm{min}}^2 + 1. \label{add_err}
\end{equation}
Here, the renormalized chi-square $\tilde{\chi}^2$ is defined as $\tilde{\chi}^2 \equiv \chi^2 / (\chi^2_{\mathrm{min}}/\nu)$ where $\chi^2_{\mathrm{min}}$ and $\nu$ are the minimum chi-square value and the number degrees of freedom in the fit analysis.
Some excited states are discussed individually below.

%

\subsubsection{The 10.996 and 11.08-MeV states}
\begin{figure}[t]
\begin{center}
 \includegraphics[width=\linewidth]{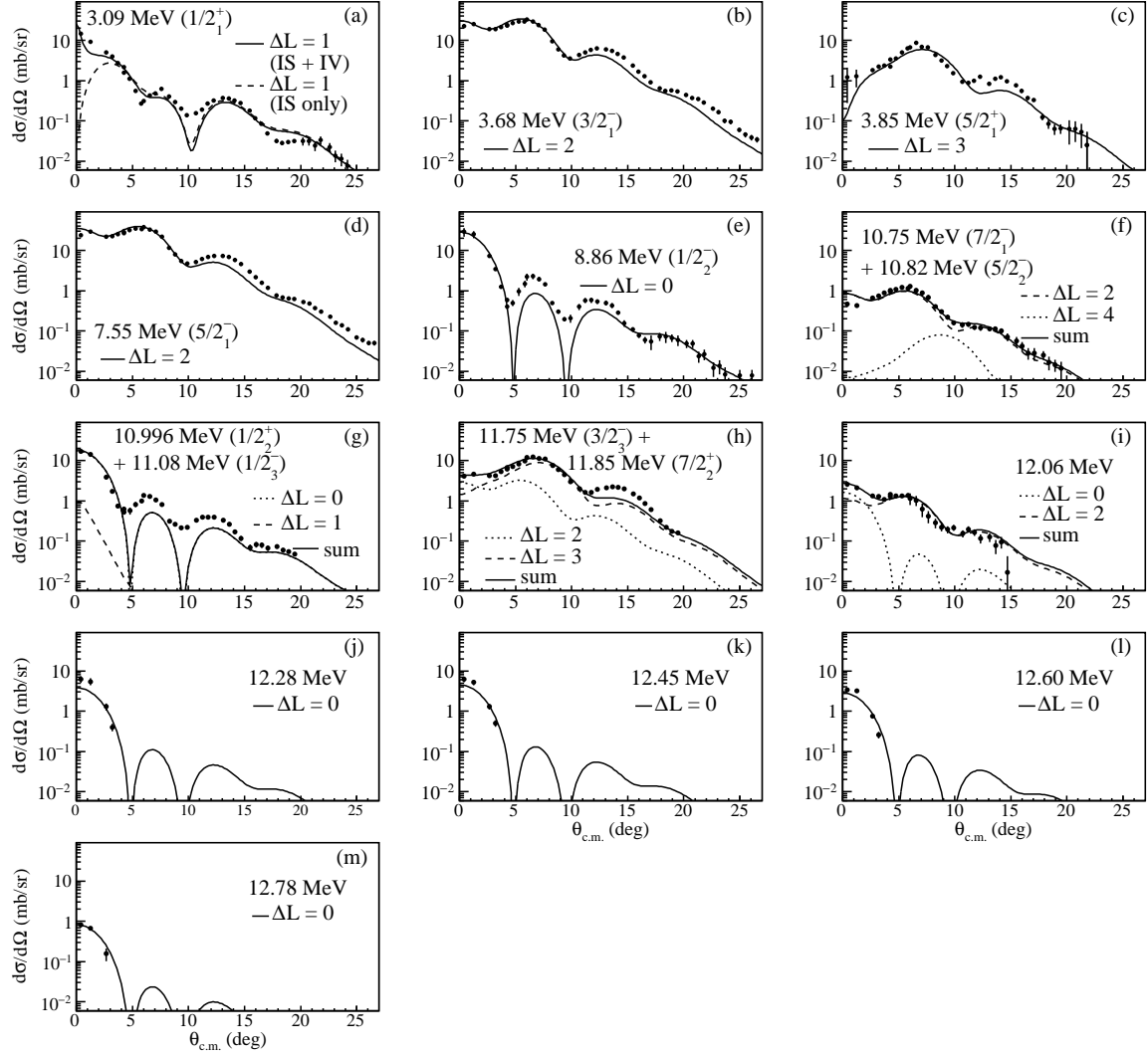}
\caption{Comparison between the DWBA cross sections and the measured cross sections in the $^{13}$C$(\alpha,\alpha')$ reaction. The solid circles show the measured cross sections. The solid, dashed, and dotted lines show the fit result with the cross sections calculated by the DWBA. In the panel (a), the solid line shows the calculated cross section considering the interference between the $IS1$ and isovector $E1$ transitions, whereas the dashed line shows the cross section for the $IS1$ transition only.}
\label{descrete_state}
\end{center} 
\end{figure}

The $1/2^+$ state at $E_x = 10.996$~MeV and the $1/2^-$ state at $E_x = 11.08$~MeV are listed in Ref.~\cite{Ajzenbergselove1991}.
The 10.996-MeV state was less populated than the 11.08-MeV state in the present experiment.
However, we included both states in the peak-fit analysis to reproduce the excitation-energy spectra.
The experimental summed cross section was fitted by the DWBA cross sections for the $\Delta L = 0$ and 1 transitions as shown in Fig.~\ref{descrete_state}(g).
The measured cross section is dominated by the $IS0$ transition, and the dotted line for the $\Delta L = 0$ transition almost overlaps with the solid line for the summed cross section.
The $E1$ strength for the 10.996-MeV state was taken from the photonuclear experiment~\cite{Jury1979}, assuming that the 10.996-MeV state corresponded to a peak structure observed at $E_x = 11.0$~MeV in Ref.~\cite{Jury1979}.  
Since the $IS1$ strength in the present DWBA analysis was zero within its uncertainty, an upper limit of the 95\% confidence interval for the $IS1$ strength was evaluated.
The spin and parity of the 11.08-MeV state are unambiguously $1/2^-$ although Milin and von Oertzen proposed that this state should be the $3/2^+$ state as the band head of the $K=3/2^+$ rotational band~\cite{Milin2002}.


\subsubsection{The 12.14-MeV state}
The $1/2^+$ state at $E_x = 12.14$~MeV is listed in Ref.~\cite{Ajzenbergselove1991}.
However, this state was not identified in the present experiment.
An upper limit of the 95\% confidence interval for the $IS1$ strength was evaluated.
The interference between the $E1$ and $IS1$ transitions was neglected in the DWBA calculation since no peak structure corresponding to the 12.14-MeV state was observed in the photonuclear experiment~\cite{Jury1979}.
This state with a broad width of 430~keV has never been clearly observed except in the $^9$Be$(\alpha, n)$ reaction~\cite{PhysRevC.49.1205, PhysRevC.53.2486}.

\subsubsection{The bump structure around $E_x =$ 12.5 MeV}
We could not reproduce the observed bump structure around $E_x = 12.5$ MeV with a single Voigtian function. We needed to introduce five new states at least in the peak-fit analysis. The centroid energies of the five new states were $E_x = 12.06$, 12.28, 12.45, 12.60, and 12.78 MeV.
These energies agreed with the centroids of the observed structures in Fig.~\ref{spectrum_fitting}.
When we added the sixth and more states in the peak-fit analysis, the chi-square of the fit was improved.
However, no correspondence between the additional states and the visible structures was observed. Therefore, we concluded that these additional states were spurious.

The cross sections for the four states at $E_x = 12.28$, 12.45, 12.60, and 12.78 MeV were well fitted by the DWBA cross sections for the $\Delta L = 0$ transition as shown in Figs.~\ref{descrete_state}(j), (k), (l), and (m), respectively.
Therefore, the spin and parity of these states were assigned to be $1/2^-$.
On the other hand, it was necessary to consider both the $\Delta L = 0$ and 2 transitions to reproduce the cross section for the 12.06-MeV state as seen in Fig.~\ref{descrete_state}(i).
This indicates that a $1/2^-$ state and a $3/2^-$ or $5/2^-$ state exist around $E_x = 12.06$ MeV although these states were not separated in the peak-fit analysis.
The energy spectrum at this energy region is so complicated that the accurate peak assignment is not easy.
However, we can conclude that several $1/2^-$ states exist around $E_x = 12.5$~MeV from the result of the peak fit and the DWBA analysis.

\begin{table}[t]
\begin{threeparttable}
\caption{$B(IS\lambda)$ strengths and EWSR fractions for the low-lying states of $^{13}$C determined by the DWBA analysis.}
 \label{strength}
\centering
\renewcommand{\arraystretch}{1.1}
 \begin{tabularx}{\linewidth}{p{25mm}p{25mm}p{30mm}p{40mm}p{40mm}} \hline \hline
\multicolumn{1}{c}{$J^\pi$} & \multicolumn{1}{c}{$E_x$} & \multicolumn{1}{c}{Multipolarity} & \multicolumn{1}{c}{$B(IS\lambda)$} & \multicolumn{1}{c}{EWSR fraction} \\
                  & \multicolumn{1}{c}{(MeV)}  & & \multicolumn{1}{c}{(fm$^{2\lambda}$)\tnote{a}} & \multicolumn{1}{c}{(\%)} \\ \hline
\multicolumn{1}{c}{$1/2_1^+$} & \multicolumn{1}{c}{3.09} &\multicolumn{1}{c}{$IS1$} & \multicolumn{1}{c}{$9.6\pm0.5$} & \multicolumn{1}{c}{$0.63\pm0.03$} \\ 
\multicolumn{1}{c}{$3/2_1^-$} & \multicolumn{1}{c}{3.68} &\multicolumn{1}{c}{$IS2$} & \multicolumn{1}{c}{$56.0\pm2.0$} & \multicolumn{1}{c}{$3.12\pm0.11$}   \\
\multicolumn{1}{c}{$5/2_1^+$} & \multicolumn{1}{c}{3.85} &\multicolumn{1}{c}{$IS3$} & \multicolumn{1}{c}{$472\pm 26$} & \multicolumn{1}{c}{$1.01\pm0.06$} \\
\multicolumn{1}{c}{$5/2_1^-$} & \multicolumn{1}{c}{7.55} &\multicolumn{1}{c}{$IS2$} & \multicolumn{1}{c}{$66.3\pm2.4$} & \multicolumn{1}{c}{$7.56\pm0.27$}    \\
\multicolumn{1}{c}{$1/2_2^-$} & \multicolumn{1}{c}{8.86} &\multicolumn{1}{c}{$IS0$} & \multicolumn{1}{c}{$29.6\pm2.3$} & \multicolumn{1}{c}{$3.95\pm0.31$} \\ \hline \hline 
\end{tabularx}
\begin{tablenotes}
 \item[a] The units of $B(IS0)$ and $B(IS1)$ are fm$^4$ and fm$^6$, respectively.
\end{tablenotes}
\end{threeparttable}
\end{table}

\subsection{Multipole decomposition analysis}
\begin{figure}[b]
 \begin{center}
 \includegraphics[width=0.7\linewidth]{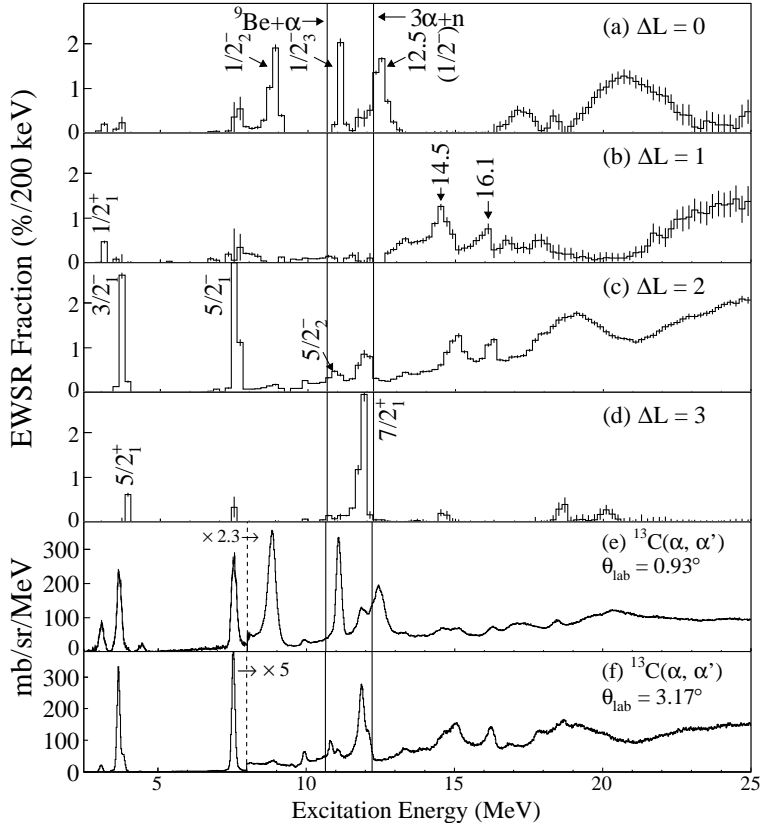}
\caption{Strength distributions for the isoscalar $\Delta L = 0$--3 transitions in the $^{13}$C$(\alpha, \alpha')$ reaction obtained by the MDA. The threshold energies for the $^9$Be$+\alpha$ and $3\alpha+n$ channels are shown by the vertical lines. The excitation-energy spectra measured at $\theta_{\mathrm{lab}} = 0.93^\circ$ and $3.17^\circ$ are also shown in the panels (e) and (f) for comparison. The spectra at $E_x > 8.0$ MeV are scaled by factors of 2.3 and 5 in the panels (e) and (f), respectively.}
\label{MDA}
 \end{center}
\end{figure}

The cross sections for the $\alpha$ inelastic scattering have characteristic angular distributions depending on the transferred angular momenta.
Therefore, the observed cross sections $\left[{d\sigma/d\Omega dE_x}\right]^{\mathrm{exp}}$ can be decomposed into each multipole component by the MDA as
\begin{equation}
 \left[ \frac{d\sigma}{d\Omega dE_x} \right]^{\mathrm{exp}} = \sum_{\lambda,~\lambda \neq 1}^{\lambda_{\mathrm{max}}} a_{\lambda}(E_x) \left[ \frac{d\sigma}{d\Omega dE_x} \right]_{IS\lambda}^{\mathrm{DWBA}} + \left[ \frac{d\sigma(\beta_1)}{d\Omega dE_x} \right]_{IS1 + E1}^{\mathrm{DWBA}}.
\end{equation}
$\left[ d\sigma/d\Omega dE_x \right]_{IS\lambda}^{\mathrm{DWBA}}$ is the calculated cross section for the $\Delta L = \lambda$ transition which exhausts 100\% of the EWSR strength, whereas $\left[ d\sigma/d\Omega dE_x \right]_{IS1+E1}^{\mathrm{DWBA}}$ is the cross section for the $\Delta L = 1$ transition taking into account the interference between the $IS1$ and isovector $E1$ transitions.
$a_\lambda(E_x)$ is a parameter corresponding to the EWSR fraction for the $\Delta L = \lambda$ transition.
$\beta_1$ denotes the collective coupling parameter for the $IS1$ transition in Eq.~\eqref{ts_1}, whose square is proportional to $B(IS1)$.
We determined $a_\lambda(E_x)$ and $\beta_1$ to minimize $\chi^2$ in the MDA under the constraint of $a_\lambda(E_x) \ge 0$.

The MDA was performed for every 200-keV excitation-energy bin up to $E_x = 25.0~\mathrm{MeV}$.
The interference between the $E1$ and $IS1$ transitions was accounted for energy bins centered at $E_x = 3.1$ MeV and at $E_x \ge 7.5$~MeV, where the radiative-decay width and the photonuclear cross section were available.
The maximum transferred angular momentum for the MDA was determined to be $\lambda_{\mathrm{max}} = 8$ because the $\chi^2/\nu$ value was almost converged at $\lambda_{\mathrm{max}} \ge 8$.

Figure \ref{MDA} shows the strength distributions for the $\Delta L = 0$--3 transitions compared with the excitation-energy spectra at $\theta_{\mathrm{lab}} = 0.93^\circ$ and $3.17^\circ$. The $IS0$ transition is dominant at $\theta_{\mathrm{lab}} = 0.93^\circ$, whereas the other transitions are stronger than the $IS0$ transition at $\theta_{\mathrm{lab}} = 3.17^\circ$.
The uncertainties of the transition strengths were estimated from the 68\% confidence intervals for $\alpha_\lambda(E_x)$ and $\beta_1$, which were determined by using Eq.~\eqref{add_err}.
When the confidence interval for one parameter was evaluated, the other parameters were freely changed to minimize $\chi^2$.

Since the spin and parity of the ground state in $^{13}$C are $1/2^-$, the spin and parity of a state excited by a $\Delta L = \lambda$ transition are $J = |1/2\pm\lambda|$ and $\pi = (-)^{(\lambda + 1)}$.
It should be noted that the known discrete states labeled in Fig.~\ref{MDA} are correctly observed in the corresponding $\Delta L = \lambda$ strength distributions.
This demonstrates the reliability of the present MDA.
Since the present DWBA calculations did not completely reproduce the cross sections for the discrete states, residues in the MDA were decomposed into incorrect strength distributions.
Therefore, spurious peaks were observed in different strength distributions near strong discrete states.
Actually, small peaks in the $\Delta L = 0, 1$, and 3 strength distributions were observed at the same energy with the $5/2_1^-$ state at $E_x = 7.55$ MeV.
These peaks were accompanied by the large uncertainties in the MDA.


\begin{figure}[t]
\begin{center}
 \includegraphics[width=\linewidth]{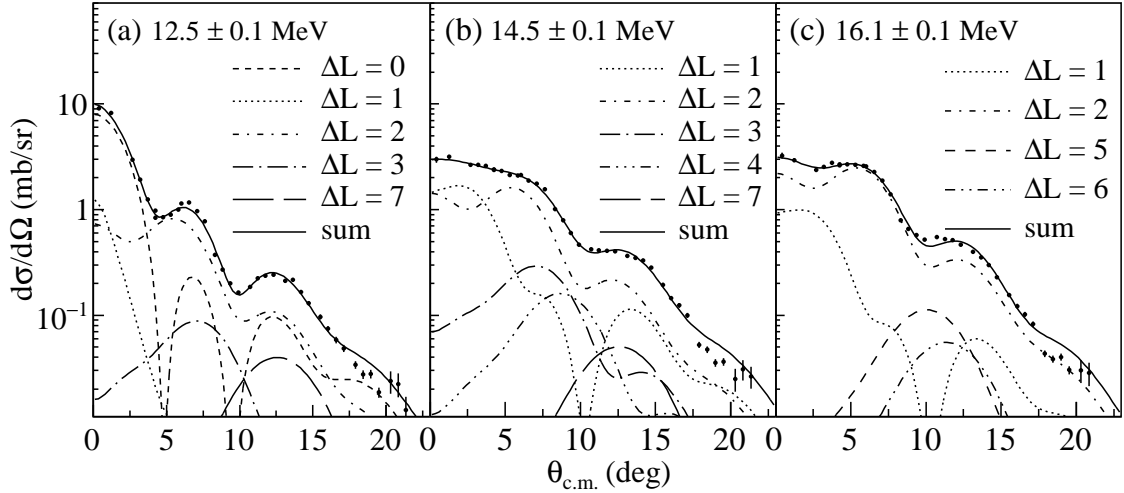}
 \caption{Measured angular distributions of the cross sections at several excitation-energy bins with 200-keV widths fitted by the MDA. The solid circles show the measured cross sections, whereas the solid lines show theoretical cross sections fitted to the experimental data. The decomposed cross sections with $a_\lambda(E_x) \neq 0$ are also plotted.}

 \label{fig:mda_distri}
\end{center} 
\end{figure}

Several narrow peaks are observed in the $\Delta L = 0$ strength distribution as shown in Fig.~\ref{MDA}(a).
In addition to the two peaks corresponding to the $1/2_2^-$ state at $E_x = 8.86$~MeV and the $1/2_3^-$ state at $E_x = 11.08$~MeV, a bump structure is observed at $E_x = 12.5$~MeV.
This bump corresponds to the $1/2^-$ states newly found around $E_x = 12.5$~MeV.
The measured angular distribution of the cross section at $E_x = 12.5$~MeV fitted by the MDA is shown in Fig.~\ref{fig:mda_distri}(a).
The steep increase of the measured cross section at forward angles is well reproduced by the $\Delta L = 0$ transition.

In the $\Delta L = 1$ strength distribution shown in Fig.~\ref{MDA}(b), not only a narrow peak due to the $1/2_1^+$ state at $E_x = 3.09$~MeV but also two bumps at $E_x = 14.5$ and 16.1~MeV are observed.
These bumps correspond to shoulders at the low-energy sides of the broad structures at $E_x = 15.0$ and 16.3 MeV in the excitation-energy spectrum at $\theta_{\mathrm{lab}} = 3.17^\circ$ presented in Fig.~\ref{MDA}(f).
The measured angular distributions of the cross sections for the excitation-energy bins at $E_x = 14.5$ and 16.1~MeV are shown in Figs.~\ref{fig:mda_distri}(b) and (c), respectively.
The sizable contributions of the $IS1$ transition to the cross sections are found.
The asymmetric shape of the bump at $E_x = 16.1$ MeV in the $\Delta L = 1$ strength distribution is due to the error of the present MDA. Since a visible peak exists around $E_x = 16.3$ MeV in the $\Delta L = 2$ strength distribution, the obtained $\Delta L = 1$ strength for the 16.3-MeV bin is affected by this peak.

\section{Discussion}
\subsection{$1/2^-$ states}
\begin{figure}[b]
 \begin{center}
 \includegraphics[width=\linewidth]{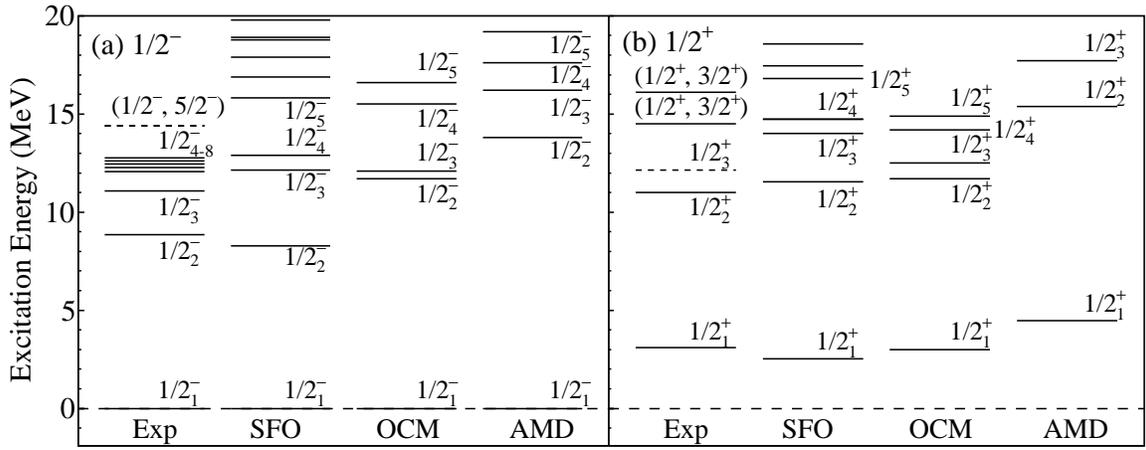}
\caption{Experimental level diagrams for (a) the $1/2^-$ states and (b) the $1/2^+$ states in $^{13}$C compared with the theoretical predictions by the shell model with the SFO interaction~\cite{Suzuki2003}, OCM~\cite{Yamada2015}, and AMD~\cite{Chiba2020}.}
\label{comp_theo}
 \end{center}
\end{figure}

\begin{figure}[t]
 \begin{center}
 \includegraphics[width=0.7\linewidth]{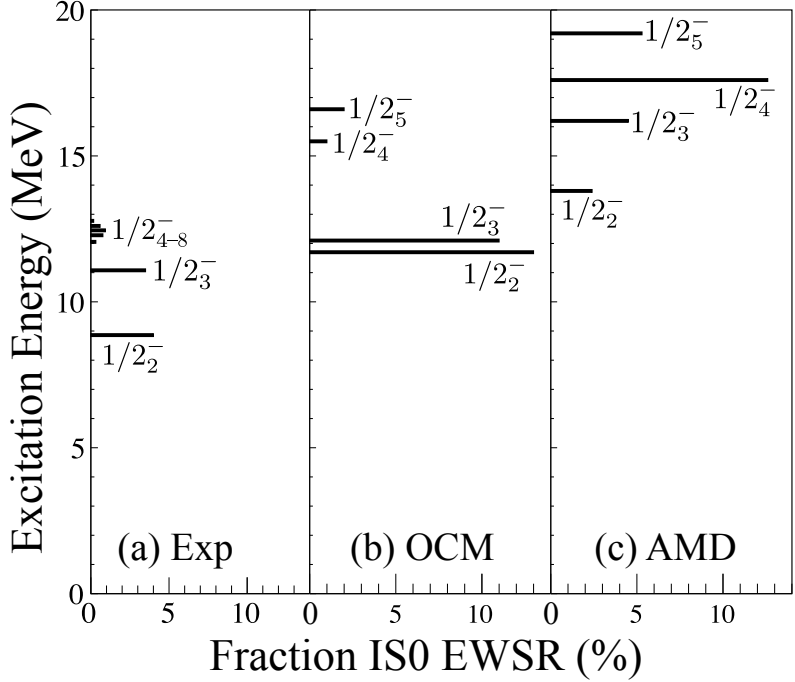}
  \caption{Measured EWSR fractions for the $IS0$ transitions compared with those predicted by the OCM~\cite{Yamada2015} and AMD~\cite{Chiba2020} calculations.}
  \label{strength_IS0}
 \end{center}
\end{figure}

The experimental level diagram for the $1/2^-$ states is compared with those obtained by several theoretical models in Fig.~\ref{comp_theo}(a).
According to the experimental compilation~\cite{Ajzenbergselove1991}, there exists a state at $E_x = 14.39$ MeV with the spin and parity of $1/2^-$ or $5/2^-$.
Since this state was not observed in the present measurement, this state is presented by the dashed line.
Although the number of the $1/2^-$ states observed around $E_x = 12.5$ MeV is more than the theoretical predictions, the shell-model calculation with the SFO interaction~\cite{Suzuki2003} best reproduces the experimental level diagram among these theoretical models.
This shell-model calculation was carried out in the $psd$ configuration space up to $2\hbar \omega$ using the code NuShellX~\cite{NuShellX}.
However, the shell model could not reproduce the experimental $IS0$ strengths.
The experimental EWSR fractions for the $IS0$ transitions are shown in Fig.~\ref{strength_IS0}(a).
The measured EWSR fractions are at least about 0.2\%, however the theoretical values calculated by the shell model are about $10^{-4}\%$.

The $3\alpha + n$~OCM~\cite{Yamada2015} and AMD~\cite{Chiba2020} calculations systematically overestimate the energies of the $1/2^-$ states.
They do not satisfactorily reproduce the observed energy levels.
However, it should be noted that the OCM and AMD calculations reproduce the sizable EWSR fractions for the $IS0$ strengths as shown in Figs.~\ref{strength_IS0}(b) and (c).
The OCM calculation explicitly takes into account the clustering degrees of freedom.
The AMD can also incorporate them.
However, the conventional shell-model calculation cannot describe the clustering degrees of freedom due to the limitation of the configuration space.
The clustering degrees of freedom are crucial to describe the sizable $IS0$ strengths for these $1/2^-$ states although the shell-model-like configurations are also needed to reproduce their energies. 
Therefore, theoretical calculations covering both the cluster-model and shell-model configuration spaces are necessary, and such calculations might solve the situation that the number of the predicted $1/2^-$ states is fewer than the experiment.

It is worth discussing the $1/2^-$ states in $^{13}$N because a good mirror symmetry is conserved between $^{13}$C and $^{13}$N.
The two $1/2^-$ states as the isobaric mirror states for the $1/2_2^-$ and $1/2_3^-$ states in $^{13}$C were clearly observed in the $^{13}$C($^{3}$He,~$t$)$^{13}$N reaction at $\theta_{\mathrm{lab}} = 0^\circ$~\cite{Fujimura2004}.
In addition to these two $1/2^-$ states, a small bump at $E_x = 13.5$ MeV was observed at forward angles.
Since this bump is considered to be excited by the Gamow-Teller transition, its spin and parity are $1/2^-$ or $3/2^-$.
It is naturally expected that the mirror state for this 13.5-MeV state is observed in the $^{13}$C($\alpha,\alpha'$) reaction.
The $1/2^-$ states observed around $E_x = 12.5$ MeV in $^{13}$C are inferred to be the mirror state of the 13.5-MeV state in $^{13}$N.

The proton decay from the excited states in $^{13}$N to the low-lying $T=0$ states in $^{12}$C were also measured in Ref.~\cite{Fujimura2004}.
The 13.5-MeV state in $^{13}$N dominantly decayed to the $0_2^+$ state in $^{12}$C as seen in Fig. 8 of Ref.~\cite{Fujimura2004} although the numerical value of the branching ratio was not given in Ref.~\cite{Fujimura2004}. Since the $1/2^-$ states around $E_x = 12.5$ MeV in $^{13}$C locate slightly above the $3\alpha + n$ decay threshold and its mirror state has large decay width to the $\alpha$ condensed state in $^{12}$C, these $1/2^-$ states in $^{13}$C might be the candidates for the $\alpha$ condensed state in which an excess neutron in the $p_{1/2}$ orbit is coupled to the $0_2^+$ state in $^{12}$C. However, the OCM calculation suggests that the wave functions of the $1/2_{2,3,4,5}^-$ states are dominated by the $^9\mathrm{Be}+\alpha$ configuration rather than the $^{12}$C$(0_2^+) + n$ configuration~\cite{Yamada2015}. Since the odd-parity $\alpha$--$n$ force is attractive~\cite{Kanada1979}, the excess neutron in the $p_{1/2}$ orbit glues two $\alpha$ clusters among the three $\alpha$ clusters in the $0_2^+$ state to form the $^9\mathrm{Be}+\alpha$ structure in the $1/2^-$ states.
Therefore, the three $\alpha$ clusters are not fully condensed into the same $S$ orbit, and the $\alpha$ condensed state is unlikely to emerge as the negative parity states.

The dilute structure of the $\alpha$ condensed state may be reflected in the isotopic shift of the excitation energies since the radial expansion of the proton distribution causes the reduction of the Coulomb energy. However, the 13.5-MeV state in $^{13}$N locates at higher excitation energy than the $1/2^-$ states around $E_x = 12.5$ MeV in $^{13}$C. This fact also suggests that these mirror states in $^{13}$N and $^{13}$C can not be regarded as the $\alpha$ condensed state.

\subsection{Positive-parity states excited by $\Delta L = 1$ transitions}
The experimental level diagram for the positive-parity states excited by the $\Delta L = 1$ transitions is compared with theoretical predictions in Fig.~\ref{comp_theo}(b).
Since the $1/2_3^+$ state at $E_x = 12.14$ MeV listed in Ref.~\cite{Ajzenbergselove1991} was not observed in this experiment, this state is presented by the dashed line.
The two states at $E_x = 14.5$ and 16.1 MeV are observed in the $\Delta L = 1$ strength distribution in Fig.~\ref{MDA}(b).
The spin and parity of these states are either $1/2^+$ or $3/2^+$.
If we assume that the spin and parity of both the 14.5-MeV and 16.1-MeV states are $1/2^+$, the theoretical level diagrams predicted by the shell-model and OCM calculations reasonably well reproduce the experiment except that the experimental $1/2_3^+$ state does not appear in the shell-model calculation.
The AMD calculation also reproduces the excitation energy of the $1/2_1^+$ state, however, the excitation energies of the $1/2_2^+$ and $1/2_3^+$ states predicted in the AMD calculation are much larger than those in the shell-model and the OCM calculations.

\begin{figure}[t]
 \begin{center}
  \includegraphics[width=\linewidth]{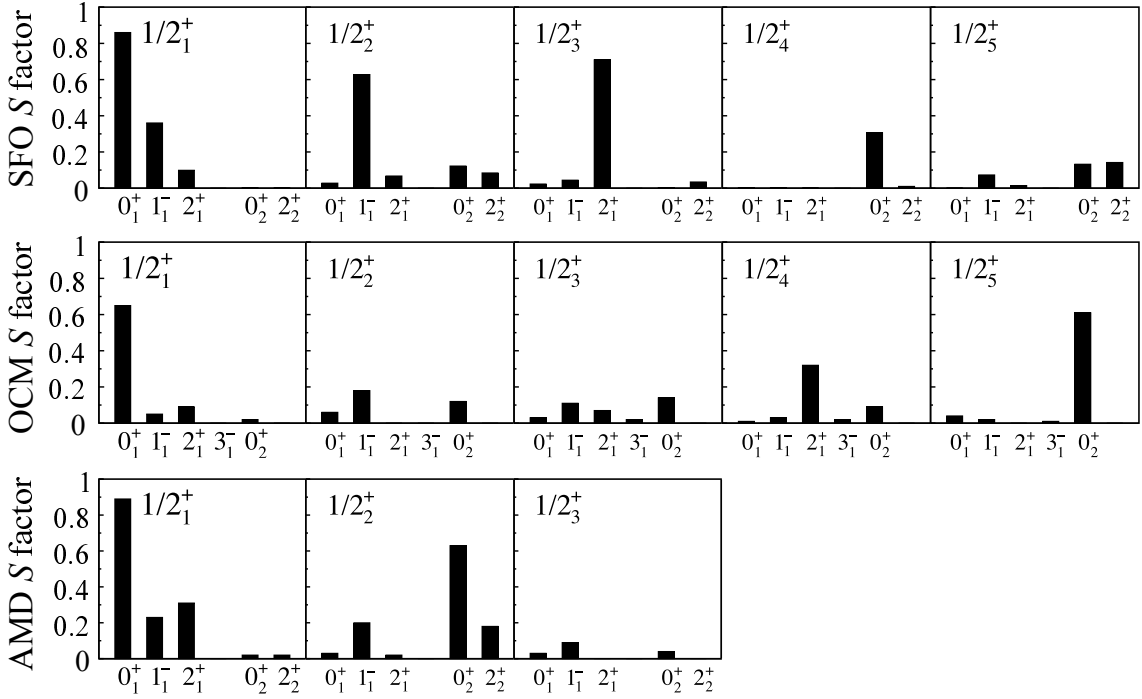}
  \caption{$S$ factors for the $^{12}\mathrm{C} + n$ channels of the $1/2^+$ states calculated by the shell model with the SFO interaction~\cite{Suzuki2003}, OCM~\cite{Yamada2015}, and AMD~\cite{Chiba2020, privatecomu}. The $^{12}\mathrm{C}(3_1^-) + n$ channel was not included in the AMD calculation, whereas the $^{12}\mathrm{C}(2_2^+) + n$ channel was not taken into account in the OCM calculation.}
  \label{sfacL1}
 \end{center}
\end{figure}

The $S$ factors of the $1/2^+$ states for the $^{12}\mathrm{C} + n$ channels calculated by the shell model, OCM, and AMD are compared in Fig.~\ref{sfacL1}.
Since the configuration space of the present shell-model calculation with the SFO interaction was limited to the $psd$ shells, the shell model did not include the $^{12}\mathrm{C}(3_1^-) + n$ channel, in which an $F$-wave neutron must be taken into account.
The AMD calculation did not include this channel either.
The $^{12}\mathrm{C}(3_1^-) + n$ channel will make a minor contribution to the $1/2^+$ states because the $S$ factor for this channel is negligibly small in the OCM calculation.

All the calculations agree that the $1/2_1^+$ state has a shell-model-like structure in which an $s_{1/2}$ neutron couples to the $0_1^+$ state in $^{12}$C.
Yamada and Funaki pointed out that the $1/2_5^+$ state in the OCM calculation is the theoretical candidate for the $\alpha$ condensed state~\cite{Yamada2015}.
The $1/2_2^+$ state in the AMD calculation also has the largest $S$ factor for the $^{12}\mathrm{C}(0_2^+) + n$ channel.
The energy of the $1/2_2^+$ state in the AMD calculation is close to that of the $1/2_5^+$ state in the OCM calculation.
Therefore, we suggest that the $1/2_{2,3,4}^+$ states in the OCM calculation are missed in the AMD calculation, and the $1/2_2^+$ state in the AMD calculation corresponds to the $1/2_5^+$ state in the OCM calculation.
Actually, Chiba and Kimura pointed out the $1/2_2^+$ state in the AMD calculation is the candidate for the $\alpha$ condensed state~\cite{Chiba2020}.

It is worth mentioning that there is a good agreement in the dominant channel between the shell-model and the OCM calculations if the $1/2_3^+$ state in the OCM calculation is ignored.
The $1/2_3^+$ state in the shell-model calculation and the $1/2_4^+$ state in the OCM calculation have the largest $S$ factor in the $^{12}\mathrm{C}(2_1^+) + n$ channel.
The $1/2_4^+$ state in the shell-model calculation and the $1/2_5^+$ state in the OCM calculation have the largest $S$ factor in the $^{12}\mathrm{C}(0_2^+) + n$ channel.
The energies of the $1/2_3^+$ and $1/2_4^+$ states in the shell-model calculation are close to those of the $1/2_4^+$ and $1/2_5^+$ states in the OCM calculation.
These states might correspond to the 14.5-MeV and 16.1-MeV states observed in the $\Delta L = 1$ strength distribution.
Therefore, the 16.1-MeV state might be the experimental candidate for the $\alpha$ condensed state.

\begin{figure}[t]
 \begin{center}
  \includegraphics[width=\linewidth]{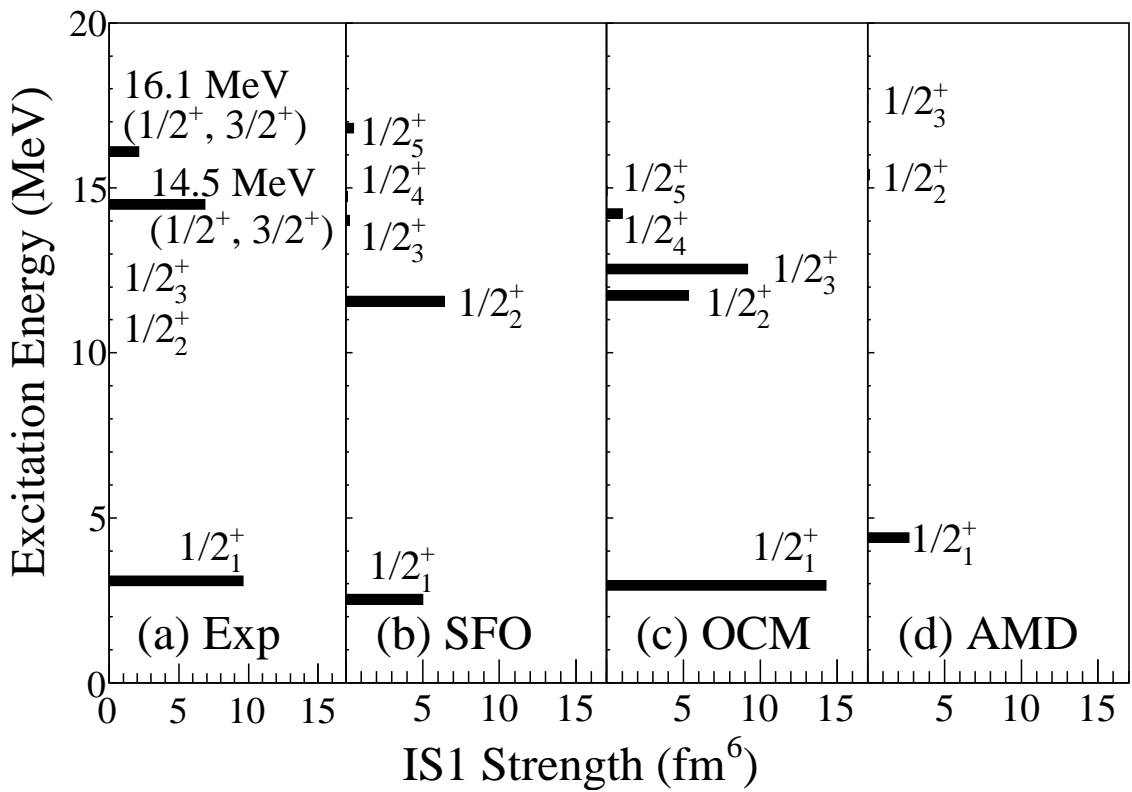}
  \caption{Measured $IS1$ strengths and those predicted by the shell model with the SFO interaction~\cite{Suzuki2003}, OCM~\cite{Yamada2015}, and AMD~\cite{Chiba2020}.}
  \label{IS1}
 \end{center}
\end{figure}

The experimental $IS1$ strengths are presented in Fig.~\ref{IS1}(a).
The $IS1$ strengths for the 14.5-MeV and 16.1-MeV states were evaluated by subtracting a continuum component of the $\Delta L = 1$ strengths in Fig.~\ref{MDA}(b) to be $6.9\pm0.7$ fm$^6$ and $2.1\pm0.8$ fm$^6$, respectively.
The theoretical $IS1$ strengths calculated by the shell model, OCM, and AMD are compared with the experimental values in Fig.~\ref{IS1}.
It should be noted that the $IS1$ strength distribution calculated by the shell model and the OCM is similar if the $1/2_3^+$ state in the OCM calculation is ignored as in the level diagram and the $S$ factors.

The measured $IS1$ strength for the $1/2_1^+$ state is reasonably well reproduced in all the theoretical calculations.
However, the sizable $IS1$ strengths for the $1/2_2^+$ state predicted by the shell-model and OCM calculations do not agree with the experiment.
An experimental counterpart of the $1/2_3^+$ state in the OCM calculation was not identified in the present study.
Although we have suggested the 16.1-MeV state is a possible candidate for the $\alpha$ condensed state on the basis of the observed level structure, the measured $IS1$ strength of this experimental candidate is much larger than the theoretical predictions.
One plausible explanation for the suppressed $IS1$ strengths in the theoretical calculations is that both of the core excitation from the $0_1^+$ state to the $0_2^+$ state in $^{12}$C and the neutron excitation from the $p_{1/2}$ orbit to the $s_{1/2}$ orbit are necessary to excite the $\alpha$ condensed state from the ground state in $^{13}$C.

Further experimental information to establish the $\alpha$ condensed state is strongly desired.
The spin and parity of the 14.5-MeV and 16.1-MeV states need to be determined.
The decay modes of these states should also be measured because the $\alpha$ condensed state is expected to decay by emitting a neutron via the $0_2^+$ state in $^{12}$C.

\section{Summary and conclusion}

We measured the $^{13}\mathrm{C}(\alpha, \alpha')$ reaction at $E_\alpha = 388$ MeV at forward angles including 0 degrees to search for the $\alpha$ condensed state in $^{13}$C.

We found a bump structure around $E_x = 12.5$ MeV due to the $IS0$ transition.
The peak-fit analysis suggested that this bump consisted of several $1/2^-$ states.
We performed the DWBA analysis with the single-folding potentials for the measured cross sections to determine the isoscalar transition strengths in $^{13}$C.
The measured angular distributions of the cross sections were reasonably well reproduced by the DWBA calculations.

We performed the MDA in order to determine the strength distributions for the isoscalar $\Delta L = 0$--3 transitions.
The known discrete states were correctly observed in the corresponding $\Delta L = \lambda$ strength distributions.
We found the two bumps at $E_x = 14.5$ and 16.1 MeV in the $\Delta L = 1$ strength distribution.
The spin and parities of these states were assigned to be $1/2^+$ or $3/2^+$.

We compared the experimental level diagram and the $IS0$ strengths for the $1/2^-$ states with those predicted by the shell model with the SFO interaction~\cite{Suzuki2003}, OCM~\cite{Yamada2015}, and AMD~\cite{Chiba2020}.
The shell-model calculation best reproduces the experimental level diagram among these calculations, but it cannot explain the sizable $IS0$ strengths for the $1/2^-$ states.
The clustering degrees of freedom are crucial to account for the sizable $IS0$ strengths for these states.
Therefore, the theoretical calculations covering both the cluster-model and shell-model configuration spaces are necessary to well describe these $1/2^-$ states.
The small bump observed around $E_x = 13.5$ MeV in $^{13}$N, which is the possible mirror states of the $1/2^-$ states around $E_x = 12.5$ MeV near the $3\alpha + n$ decay threshold in $^{13}$C, dominantly decays to the $0_2^+$ state in $^{12}$C~\cite{Fujimura2004}.
This fact implies that the $1/2^-$ states around $E_x = 12.5$ MeV is the candidate for the $\alpha$ condensed state in which a $p_{1/2}$ neutron couples to the $^{12}$C$(0_2^+)$ core.
However, the OCM calculation suggests that the wave functions of the $1/2_{2,3,4,5}^-$ states are dominated by the $^9\mathrm{Be} + \alpha$ configuration due to the attractive odd-parity $\alpha$--$n$ force, and thus the $\alpha$ condensed state is unlikely to emerge as the negative parity states in $^{13}$C.

The experimental level diagram for the positive parity states excited with the $IS1$ strength is reasonably well reproduced by the shell-model and OCM calculations if we assume the spin and parity of both the 14.5-MeV and 16.1-MeV states to be $1/2^+$.
On the other hand, the excitation energies of the $1/2_2^+$ and $1/2_3^+$ states predicted in the AMD calculation are much larger than those in the shell-model and the OCM calculations.
We suggest that the AMD calculation misses the $1/2_{2,3,4}^+$ states predicted by the OCM calculation.
All the calculations reasonably well reproduce the measured $IS1$ strength for the $1/2_1^+$ state, which is suggested to have a shell-model-like structure in these calculations.
We suggest that the 16.1-MeV state is a possible candidate for the $\alpha$ condensed state predicted by the OCM and AMD calculations on the basis of the good correspondence between the experimental and calculated level structures.
However, the measured $IS1$ strength for the 16.1-MeV state is much larger than the theoretical predictions.
Therefore, further experimental information is required to establish the $\alpha$ condensed state in $^{13}$C.
The spin and parity of the 14.5-MeV and 16.1-MeV states need to be determined.
The decay modes of these states should also be investigated since the $\alpha$ condensed state is expected to decay by emitting a neutron via the $0_2^+$ state in $^{12}$C.
A new experiment measuring the decay neutrons from the 14.5-MeV and 16.1-MeV states in coincidence with inelastically scattered $\alpha$ particles is highly desired.

\section*{Acknowledgements}
The authors are grateful to the RCNP cyclotron crews for the stable operation of the cyclotron facilities. 
The authors also thank Dr. Y. Chiba from Osaka City University and Prof. M. Kimura from Hokkaido University for the fruitful discussion on the AMD calculation for $^{13}$C, and Prof. I. Sugai for preparing the high quality $^{\mathrm{nat}}$C and $^{13}$C targets. 
K. I. appreciates the support of Grant-in-Aid for JSPS Research Fellow JP20J15126.
This research was supported by JSPS KAKENHI Grant Number JP17740132.


\begin{thebibliography}{10}

\bibitem{Horiuchi1968}
H.~Horiuchi, K.~Ikeda, and N.~Takigawa, Prog. Theor. Phys. Suppl. {\bf E68},
  464 (1968).

\bibitem{Fujiwara1966}
Y.~Fujiwara, H.~Horiuchi, K.~Ikeda, M.~Kamimura, K.~Kato, Y.~Suzuki, and
  E.~Uegaki, Prog. Theor. Phys. Suppl. {\bf 68}, 29 (1966).

\bibitem{Uegaki1977}
E.~Uegaki, S.~Okabe, Y.~Abe, and H.~Tanaka, Prog. Theor. Phys. {\bf 57}, 1262
  (1977).

\bibitem{Kamimura1981}
M.~Kamimura, Nucl. Phys. A {\bf 351}, 456 (1981).

\bibitem{Navratil2003}
P.~Navr\'{a}til and W.~E. Ormand, Phys. Rev. C {\bf 68}, 034305 (2003).

\bibitem{Morinaga1956}
H.~Morinaga, Phys. Rev. {\bf 101}, 254 (1956).

\bibitem{Morinaga1966}
H.~Morinaga, Phys. Lett. {\bf 21}, 78 (1966).

\bibitem{Tohsaki2001}
A.~Tohsaki, H.~Horiuchi, P.~Schuck, and G.~R\"opke, Phys. Rev. Lett. {\bf 87},
  192501 (2001).

\bibitem{Funaki2003}
Y.~Funaki, A.~Tohsaki, H.~Horiuchi, P.~Schuck, and G.~R{\"{o}}pke, Phys. Rev.
  C {\bf 67}, 051306 (2003).

\bibitem{Yamada2005}
T.~Yamada and P.~Schuck, Euro. Phys. J. A {\bf 26}, 185 (2005).

\bibitem{Yamada2004}
T.~Yamada and P.~Schuck, Phys. Rev. C {\bf 69}, 024309 (2004).

\bibitem{Wakasa2007}
T.~Wakasa, E.~Ihara, K.~Fujita, Y.~Funaki, K.~Hatanaka, H.~Horiuchi, M.~Itoh,
  J.~Kamiya, G.~R{\"{o}}pke, H.~Sakaguchi, N.~Sakamoto, Y.~Sakemi, P.~Schuck,
  Y.~Shimizu, M.~Takashina, S.~Terashima, A.~Tohsaki, M.~Uchida, H.~P. Yoshida,
  and M.~Yosoi, Phys. Lett. B {\bf 653}, 173 (2007).

\bibitem{Itoh2010}
T.~Hayamizu A. Oikawa Y.~Sakemi M.~Itoh, T.~Takahashi and H.~Yoshida, Mod.
  Phys. Lett. A {\bf 25}, 1935 (2010).

\bibitem{Ogloblin2016}
A.~A. Ogloblin, A.~N. Danilov, A.~S. Demyanova, S.~A. Goncharov, and T.~L.
  Belyaeva, Phys. Rev. C {\bf 94}, 051602 (2016).

\bibitem{Li2017}
K.~C.~W. Li, R.~Neveling, P.~Adsley, P.~Papka, F.~D. Smit, J.~W. Br\"ummer,
  C.~Aa. Diget, M.~Freer, M.~N. Harakeh, Tz. Kokalova, F.~Nemulodi,
  L.~Pellegri, B.~Rebeiro, J.~A. Swartz, S.~Triambak, J.~J. van Zyl, and
  C.~Wheldon, Phys. Rev. C {\bf 95}, 031302 (2017).

\bibitem{Adachi2021}
S.~Adachi, Y.~Fujikawa, T.~Kawabata, H.~Akimune, T.~Doi, T.~Furuno, T.~Harada,
  K.~Inaba, S.~Ishida, M.~Itoh, C.~Iwamoto, N.~Kobayashi, Y.~Maeda, Y.~Matsuda,
  M.~Murata, S.~Okamoto, A.~Sakaue, R.~Sekiya, A.~Tamii, and M.~Tsumura, Phys. Lett. B {\bf 819}, 136411 (2021).

\bibitem{Kawabata2013}
T.~Kawabata, T.~Adachi, M.~Fujiwara, K.~Hatanaka, Y.~Ishiguro, M.~Itoh,
  Y.~Maeda, H.~Matsubara, H.~Miyasako, Y.~Nozawa, T.~Saito, S.~Sakaguchi,
  Y.~Sasamoto, Y.~Shimizu, T.~Takahashi, A.~Tamii, S.~Terashima, H.~Tokieda,
  N.~Tomida, T.~Uesaka, M.~Uchida, Y.~Yasuda, N.~Yokota, H.~P. Yoshida, and
  J.~Zenihiro, Few-Body Systems {\bf 54}, 93 (2013).

\bibitem{Bishop2019}
J. Bishop, Tz. Kokalova, M.~Freer, L. Acosta, M. Assié, S. Bailey, G. Cardella, N. Curtis, E. De Filippo, D. Dell' Aquila, S. De Luca, L. Francalanza, B. Gnoffo, G. Lanzalone, I. Lombardo,
  N.S. Martorana, S. Norella, A. Pagano, E.V. Pagano, M. Papa, S. Pirrone, G. Politi,
  F. Rizzo, P. Russotto, L. Quattrocchi, R. Smith, I. Stefan, A. Trifirò, M.
  Trimarchì, G. Verde, M.~Vigilante, and C.~Wheldon,
  Phys. Rev. C {\bf 100}, 034320 (2019).

\bibitem{Kawabata2007}
T.~Kawabata, H.~Akimune, H.~Fujita, Y.~Fujita, M.~Fujiwara, K.~Hara,
  K.~Hatanaka, M.~Itoh, Y.~Kanada-En'yo, S.~Kishi, K.~Nakanishi, H.~Sakaguchi,
  Y.~Shimbara, A.~Tamii, S.~Terashima, M.~Uchida, T.~Wakasa, Y.~Yasuda, H.~P.
  Yoshida, and M.~Yosoi, Phys. Lett. B {\bf 646}, 6 (2007).

\bibitem{Ajzenbergselove1991}
F.~Ajzenberg-Selove, Nucl. Phys. A {\bf 523}, 1 (1991).

\bibitem{Yamada2010}
T.~Yamada and Y.~Funaki, Phys. Rev. C {\bf 82}, 064315 (2010).

\bibitem{Kanada-EnYo2015}
Y.~Kanada-En'Yo and T.~Suhara, Phys. Rev. C {\bf 91}, 014316 (2015).

\bibitem{Zhou2018}
B.~Zhou and M.~Kimura, Phys. Rev. C {\bf 98}, 054323 (2018).

\bibitem{Yamada2015}
T.~Yamada and Y.~Funaki, Phys. Rev. C {\bf 92}, 034326 (2015).

\bibitem{Chiba2020}
Y.~Chiba and M.~Kimura, Phys. Rev. C {\bf 101}, 024317 (2020).

\bibitem{Adachi2018}
S.~Adachi, T.~Kawabata, K.~Minomo, T.~Kadoya, N.~Yokota, H.~Akimune, T.~Baba,
  H.~Fujimura, M.~Fujiwara, Y.~Funaki, T.~Furuno, T.~Hashimoto, K.~Hatanaka,
  K.~Inaba, Y.~Ishii, M.~Itoh, C.~Iwamoto, K.~Kawase, Y.~Maeda, H.~Matsubara,
  Y.~Matsuda, H.~Matsuno, T.~Morimoto, H.~Morita, M.~Murata, T.~Nanamura,
  I.~Ou, S.~Sakaguchi, Y.~Sasamoto, R.~Sawada, Y.~Shimizu, K.~Suda, A.~Tamii,
  Y.~Tameshige, M.~Tsumura, M.~Uchida, T.~Uesaka, H.~P. Yoshida, and
  S.~Yoshida, Phys. Rev. C {\bf 97}, 014601 (2018).

\bibitem{PhysRevC.33.1116}
H.~J. Lu, S.~Brandenburg, R.~De~Leo, M.~N. Harakeh, T.~D. Poelhekken, and
  A.~van~der Woude, Phys. Rev. C {\bf 33}, 1116 (1986).

\bibitem{Youngblood1999}
D.~H. Youngblood, Y.-W. Lui, and H.~L. Clark, Phys. Rev. C {\bf 60}, 014304
  (1999).

\bibitem{Itoh2002}
M.~Itoh, H.~Sakaguchi, M.~Uchida, T.~Ishikawa, T.~Kawabata, T.~Murakami,
  H.~Takeda, T.~Taki, S.~Terashima, N.~Tsukahara, Y.~Yasuda, M.~Yosoi, U.~Garg,
  M.~Hedden, B.~Kharraja, M.~Koss, B.K. Nayak, S.~Zhu, H.~Fujimura,
  M.~Fujiwara, K.~Hara, H.P. Yoshida, H.~Akimune, M.N. Harakeh, and
  M.~Volkerts, Phys. Lett. B {\bf 549}, 58 (2002).

\bibitem{Uchida2003}
M.~Uchida, H.~Sakaguchi, M.~Itoh, M.~Yosoi, T.~Kawabata, H.~Takeda, Y.~Yasuda,
  T.~Murakami, T.~Ishikawa, T.~Taki, N.~Tsukahara, S.~Terashima, U.~Garg,
  M.~Hedden, B.~Kharraja, M.~Koss, B.K. Nayak, S.~Zhu, M.~Fujiwara,
  H.~Fujimura, K.~Hara, E.~Obayashi, H.P. Yoshida, H.~Akimune, M.N. Harakeh,
  and M.~Volkerts, Phys. Lett. B {\bf 557}, 12 (2003).

\bibitem{Itoh2003}
M.~Itoh, H.~Sakaguchi, M.~Uchida, T.~Ishikawa, T.~Kawabata, T.~Murakami,
  H.~Takeda, T.~Taki, S.~Terashima, N.~Tsukahara, Y.~Yasuda, M.~Yosoi, U.~Garg,
  M.~Hedden, B.~Kharraja, M.~Koss, B.~K. Nayak, S.~Zhu, H.~Fujimura,
  M.~Fujiwara, K.~Hara, H.~P. Yoshida, H.~Akimune, M.~N. Harakeh, and
  M.~Volkerts, Phys. Rev. C {\bf 68}, 064602 (2003).

\bibitem{Uchida2004}
M.~Uchida, H.~Sakaguchi, M.~Itoh, M.~Yosoi, T.~Kawabata, Y.~Yasuda, H.~Takeda,
  T.~Murakami, S.~Terashima, S.~Kishi, U.~Garg, P.~Boutachkov, M.~Hedden,
  B.~Kharraja, M.~Koss, B.~K. Nayak, S.~Zhu, M.~Fujiwara, H.~Fujimura, H.~P.
  Yoshida, K.~Hara, H.~Akimune, and M.~N. Harakeh, Phys. Rev. C {\bf 69},
  051301 (2004).

\bibitem{Li2007}
T.~Li, U.~Garg, Y.~Liu, R.~Marks, B.~K. Nayak, P.~V.~Madhusudhana Rao,
  M.~Fujiwara, H.~Hashimoto, K.~Kawase, K.~Nakanishi, S.~Okumura, M.~Yosoi,
  M.~Itoh, M.~Ichikawa, R.~Matsuo, T.~Terazono, M.~Uchida, T.~Kawabata,
  H.~Akimune, Y.~Iwao, T.~Murakami, H.~Sakaguchi, S.~Terashima, Y.~Yasuda,
  J.~Zenihiro, and M.~N. Harakeh, Phys. Rev. Lett. {\bf 99}, 162503 (2007).

\bibitem{Itoh2011}
M.~Itoh, H.~Akimune, M.~Fujiwara, U.~Garg, N.~Hashimoto, T.~Kawabata,
  K.~Kawase, S.~Kishi, T.~Murakami, K.~Nakanishi, Y.~Nakatsugawa, B.~K. Nayak,
  S.~Okumura, H.~Sakaguchi, H.~Takeda, S.~Terashima, M.~Uchida, Y.~Yasuda,
  M.~Yosoi, and J.~Zenihiro, Phys. Rev. C {\bf 84}, 054308 (2011).

\bibitem{Itoh2013}
M.~Itoh, S.~Kishi, H.~Sakaguchi, H.~Akimune, M.~Fujiwara, U.~Garg, K.~Hara,
  H.~Hashimoto, J.~Hoffman, T.~Kawabata, K.~Kawase, T.~Murakami, K.~Nakanishi,
  B.~K. Nayak, S.~Terashima, M.~Uchida, Y.~Yasuda, and M.~Yosoi, Phys. Rev. C
  {\bf 88}, 064313 (2013).

\bibitem{Gupta2016}
Y.~K. Gupta, U.~Garg, J.~Hoffman, J.~Matta, P.~V.~Madhusudhana Rao, D.~Patel,
  T.~Peach, K.~Yoshida, M.~Itoh, M.~Fujiwara, K.~Hara, H.~Hashimoto,
  K.~Nakanishi, M.~Yosoi, H.~Sakaguchi, S.~Terashima, S.~Kishi, T.~Murakami,
  M.~Uchida, Y.~Yasuda, H.~Akimune, T.~Kawabata, and M.~N. Harakeh, Phys. Rev.
  C {\bf 93}, 044324 (2016).

\bibitem{Peach2016}
T.~Peach, U.~Garg, Y.~K. Gupta, J.~Hoffman, J.~T. Matta, D.~Patel,
  P.~V.~Madhusudhana Rao, K.~Yoshida, M.~Itoh, M.~Fujiwara, K.~Hara,
  H.~Hashimoto, K.~Nakanishi, M.~Yosoi, H.~Sakaguchi, S.~Terashima, S.~Kishi,
  T.~Murakami, M.~Uchida, Y.~Yasuda, H.~Akimune, T.~Kawabata, M.~N. Harakeh,
  and G.~Col\`o, Phys. Rev. C {\bf 93}, 064325 (2016).

\bibitem{GiantResonance}
M.~N. Harakeh and A.~van~der Woude, ``Giant Resonances'', Oxford University Press, (2001).

\bibitem{Wakasa2002}
T.~Wakasa, K.~Hatanaka, Y.~Fujita, G.P.A Berg, H.~Fujimura, H.~Fujita, M.~Itoh,
  J.~Kamiya, T.~Kawabata, K.~Nagayama, T.~Noro, H.~Sakaguchi, Y.~Shimbara, H.~Takeda,
  K.~Tamura, H.~Ueno, M.~Uchida, M.~Uraki, and M.~Yosoi, Nucl. Instrum. Methods in Phys. Res. A {\bf 482}, 79 (2002).

\bibitem{Fujiwara1999}
M.~Fujiwara, H.~Akimune, I.~Daito, H.~Fujimura, Y.~Fujita, K.~Hatanaka,
  H.~Ikegami, I.~Katayama, K.~Nagayama, N.~Matsuoka, S.~Morinobu, T.~Noro,
  M.~Yoshimura, H.~Sakaguchi, Y.~Sakemi, A.~Tamii, and M.~Yosoi, Nucl. Instrum.
  Methods Phys. Res., Sect. A {\bf 422}, 484 (1999).

\bibitem{ECIS95}
J.~Raynel, ``Computer program: ECIS-95'' (1995), old version of ECIS-12 NEA-0850/19.

\bibitem{DeVries1987}
H.~{De Vries}, C.~W. {De Jager}, and C.~{De Vries}, At. Data Nucl.
  Data Tables {\bf 36}, 495 (1987).

\bibitem{Satchler1997}
G.~R. Satchler and D.~T. Khoa, Phys. Rev. C {\bf 55}, 285 (1997).

\bibitem{Furuno2019}
T.~Furuno, T.~Kawabata, S.~Adachi, Y.~Ayyad, Y.~Kanada-En'Yo, Y.~Fujikawa,
  K.~Inaba, M.~Murata, H.~J. Ong, M.~Sferrazza, Y.~Takahashi, T.~Takeda,
  I.~Tanihata, D.~T. Tran, and M.~Tsumura, Phys. Rev. C {\bf 100}, 054322 (2019).

\bibitem{Harakeh1981}
M.~N. Harakeh and A.~E.L. Dieperink, Phys. Rev. C {\bf 23}, 2329 (1981).

\bibitem{Satchler1987}
G.~R. Satchler, Nucl. Phys. A {\bf 472}, 215 (1987).

\bibitem{Endt1993}
P.~M. Endt, At. Data Nucl. Data Tables {\bf 55}, 171 (1993).

\bibitem{Jury1979}
J.~W. Jury, B.~L. Berman, D.~D. Faul, P.~Meyer, K.~G. McNeill, and J.~G.
  Woodworth, Phys. Rev. C, {\bf 19} 1684 (1979).

\bibitem{Zubanov1983}
D.~Zubanov, R.~A. Sutton, M.~N. Thompson, and J.~W. Jury, Phys. Rev. C {\bf
  27}, 1957 (1983).

\bibitem{PhysRevC.55.285}
G.~R. Satchler and D. T. Khoa, Phys. Rev. C {\bf 55}, 285 (1997).

\bibitem{Zenihiro2010}
J.~Zenihiro, H.~Sakaguchi, T.~Murakami, M.~Yosoi, Y.~Yasuda, S.~Terashima,
  Y.~Iwao, H.~Takeda, M.~Itoh, H.~P. Yoshida, and M.~Uchida, Phys. Rev. C {\bf 82}, 044611 (2010).

\bibitem{Milin2002}
M.~Milin and W.~von Oertzen, Eur.~Phys.~J.~A {\bf 14}, 295 (2002).

\bibitem{PhysRevC.49.1205}
P.~R. Wrean, C.~R. Brune, and R.~W. Kavanagh, Phys. Rev. C, {\bf 49}
  1205 (1994).

\bibitem{PhysRevC.53.2486}
R.~Kunz, S.~Barth, A.~Denker, H.~W. Drotleff, J.~W. Hammer, H.~Knee, and
  A.~Mayer, Phys. Rev. C {\bf 53}, 2486 (1996).

\bibitem{Suzuki2003}
T. Suzuki, R. Fujimoto, and T. Otsuka, Phys. Rev. C {\bf 67}, 044302 (2003).

\bibitem{NuShellX}
B.~A. Brown and W.~D.~M. Rae, ``The Shell-Model Code NuShellX'', Nucl. Data
  Sheets {\bf 120}, 115 (2014).

\bibitem{Fujimura2004}
H.~Fujimura, H.~Akimune, I.~Daito, M.~Fujiwara, K.~Hara, K.~Y. Hara, M.~N.
  Harakeh, F.~Ihara, T.~Inomata, K.~Ishibashi, T.~Ishikawa, T.~Kawabata,
  A.~Tamii, M.~Tanaka, H.~Toyokawa, T.~Yamanaka, and M.~Yosoi, Phys. Rev.
  C {\bf 69}, 064327 (2004).

\bibitem{Kanada1979}
H.~Kanada, T.~Kaneko, S.~Nagata, and M.~Nomoto, Prog. Theor. Phys. {\bf 61},
  1327 (1979).
\bibitem{privatecomu}
Y.~Chiba (private communication).
\end{thebibliography}
%

\end{document}